\documentclass[11pt,a4paper]{article} 

\pdfoutput=1
\usepackage{jcappub}
\usepackage{times}

\usepackage{graphicx}
\usepackage{dcolumn}
\usepackage{bm}
\usepackage{color}
\usepackage{calrsfs}
\usepackage{hyperref}
\usepackage{multirow} 
\usepackage{float}
\usepackage{subfloat}
\usepackage{caption}
\usepackage{subcaption}
\usepackage{soul}

\input epsf

\newcommand{\be}{\begin{equation}}
\newcommand{\ee}{\end{equation}}
\newcommand{\bear}{\begin{eqnarray}}
\newcommand{\ear}{\end{eqnarray}}

\newcommand{\f}{\frac}
\newcommand{\de}{{\rm d}}

\renewcommand{\vec}[1]{\mathbf{#1}}

\newcommand{\lya} {Ly$\alpha$\,}

\begin{document}

\title{The cross-correlation between 21cm intensity mapping maps and the Ly$\alpha$ forest in
the post-reionization era}

\author[a,b]{Isabella P. Carucci,} 
\author[c,b,d]{Francisco Villaescusa-Navarro,}
\author[d,b]{Matteo Viel}

\affiliation[a]{SISSA- International School for Advanced Studies, Via Bonomea 265, 34136 Trieste, Italy}
\affiliation[b]{INFN sez. Trieste, Via Valerio 2, 34127 Trieste, Italy}
\affiliation[c]{Center for Computational Astrophysics, 160 5th Avenue, New York, NY, 10010, USA}
\affiliation[d]{INAF - Osservatorio Astronomico di Trieste, Via Tiepolo 11, 34143, Trieste, Italy}

\emailAdd{ipcarucci@sissa.it}
\emailAdd{fvillaescusa@simonsfoundation.org}
\emailAdd{viel@oats.inaf.it}

\abstract{We investigate the cross-correlation signal between 21cm intensity mapping maps and the Ly$\alpha$ forest in the fully non-linear regime using state-of-the-art hydrodynamic simulations. The cross-correlation signal between the Ly$\alpha$ forest and 21cm maps can provide a coherent and comprehensive picture of the neutral hydrogen (HI) content of our Universe in the post-reionization era, probing both its mass content and volume distribution. We compute the auto-power spectra of both fields together with their cross-power spectrum at $z=2.4$ and find that on large scales the fields are completely anti-correlated. This anti-correlation arises because regions with high (low) 21cm emission, such as those with a large (low) concentration of damped Ly$\alpha$ systems, will show up as regions with low (high) transmitted flux. We find that on scales smaller than $k\simeq0.2~h{\rm Mpc}^{-1}$ the cross-correlation coefficient departs from $-1$, at a scale where non-linearities show up. We use the anisotropy of the power spectra in redshift-space to determine the values of the bias and of the redshift-space distortion parameters of both fields. We find that the errors on the value of the cosmological and astrophysical parameters  could decrease by 30\% when adding data from the cross-power spectrum, in a conservative analysis. Our results point out that linear theory is capable of reproducing the shape and amplitude of the cross-power up to rather non-linear scales. Finally, we find that the 21cm-Ly$\alpha$ cross-power spectrum can be detected by combining data from a BOSS-like survey together with 21cm intensity mapping observations by SKA1-MID with a S/N ratio higher than 3 in $k\in[0.06,1]~h{\rm Mpc}^{-1}$. We emphasize that while the shape and amplitude of the 21cm auto-power spectrum can be severely affected by residual foreground contamination, cross-power spectra will be less sensitive to that and therefore can be used to identify systematics in the 21cm maps.}

\maketitle

\section{Introduction}
\label{sec:introduction}

The discovery of the existence of both dark matter and dark energy has revolutionized our understanding of the fundamental constituents of the Universe. The standard model of cosmology has been extremely successful in providing the theoretical framework that is capable of explaining a very diverse set of cosmological observations ranging from the anisotropies in the cosmic microwave background (CMB) to the spatial distribution of galaxies at late times. 

The nature of dark energy, assumed to be a cosmological constant in the standard $\Lambda$CDM model, remains a mystery. Its presence and properties leaves however a signature on the growth and spatial distribution of matter perturbations. Thus, cosmological observations of those can be used to improve our understanding on the physical nature of dark energy. 

While the spatial distribution of matter in the Universe is not directly observable, its statistical properties can be examined through its tracers such as galaxies or cosmic neutral hydrogen (HI). The underlying idea is that the clustering properties of the tracers will, up to a bias factor, resemble those of the matter perturbations on large linear scales. 

Neutral hydrogen can be detected in the Universe either in emission or in absorption. For the latter, observations of the Ly$\alpha$ forest in quasar spectra represents a powerful way to trace high-redshift ($z\sim[2-4]$) highly ionized low-density gas clouds where HI is found in absorption, while for the former the 21cm line of neutral hydrogen can be used to trace HI in emission. 21cm intensity mapping is a technique \citep{Bharadwaj_2001A, Bharadwaj_2001B, Battye:2004re,McQuinn_2006, Chang_2008,Loeb_Wyithe_2008, Bull_2015} that consists in detecting the integrated 21cm emission from cosmic neutral hydrogen from unresolved galaxies or HI clouds by means of low angular resolution radio observations. The very large volumes that can be surveyed together with the intrinsic spectroscopic nature of these observations make this technique a potential game changer in the epoch of precision cosmology. 

The total radiation a radio-telescope collects is the sum of the HI cosmological signal, system noise and foregrounds. Unfortunately, the amplitude of the foregrounds (mainly synchrotron radiation) is several orders of magnitude above the one of the cosmological signal. The potential to investigate cosmological problems depends dramatically on the precision with which we will be able to remove the foregrounds from the 21cm maps \cite{Wolz_2015}. While \textit{well behaved} foregrounds can be robustly subtracted, the presence of some foregrounds such as polarized synchrotron radiation may not be removed and can significantly bias the inferred shape and amplitude of the 21cm power spectrum \cite{Alonso_2015, villaescusa2015}. 

A way to avoid residual foreground contamination is to carry out cross-correlations: since the spatial location and amplitude of the residual foregrounds is not correlated with the cosmological location of other tracers, cross-correlations should retrieve the underlying cosmological signal \cite{villaescusa2015, Sobacchi_2016,Wolz_2016, Lidz_2009,sarkar,sarkar2015,Sarkar_2016}. Furthermore, cross-correlations are very useful because they provide extra information that can be used to tighten the constraints on bias parameters and therefore to break degeneracies with other parameters.

The goal of this paper is to investigate the signal arising from the cross-correlation between the Ly$\alpha$ forest and 21cm intensity mapping maps. We notice that this analysis has already been carried out at linear order in \cite{sarkar}, while here we investigate this cross-correlation in the fully non-linear regime by means of  hydrodynamic simulations. In this paper we study the degree and scales where the two tracers are correlated and how well linear theory is able to reproduce the results.

We carry out our analysis making use of state-of-the-art hydrodynamic simulations and perform convergence tests using simulations with different resolutions. We also forecast the precision with which the future Square Kilometre Array\footnote{https://www.skatelescope.org} will be able to detect the 21cm auto-power spectrum together with the 21cm-Ly$\alpha$ cross-power spectrum at $z\simeq2.4$.

This paper is organized as follows. We start in section \ref{sec:simulations} by describing the numerical simulations which we have analyzed. Then, in sections \ref{sec:21cm_distribution}-\ref{sec:Lya} we describe how we have modeled the observables we are interested in: the 21cm radiation and the Ly$\alpha$ forest flux. In section \ref{sec:results} we show the results of the cross-correlation procedure, performing a detectability study with present and future surveys and quantifying the bias parameters of the two fields. We summarize and draw our conclusions in section \ref{sec:conclusions}. Appendices \ref{appHI} and \ref{app} focus on the modelling of the two fields: on the way in which we distribute neutral hydrogen in halos and on the way we generate mock Ly$\alpha$ forest spectra. In appendix \ref{app_errors} we derive the analytical formulas we use to model errors in our analysis.
\section{Simulations and halo catalogues}
\label{sec:simulations}

We rely on two high-resolution hydrodynamic simulations, that we label 80-2048 and 160-1024: the first number indicates the box size in $h^{-1}$Mpc while the second one represents the cube root number of gas (and dark matter) particles. Both of them are part of  the Sherwood simulation suite of Bolton et al. \cite{bolton2016}. They are run using the TreePM+SPH code {\sc Gadget-III}, an updated and extended version of {\sc Gadget-II} \cite{Springel_2005}. The values of the cosmological parameters are in agreement with recent Planck data \citep{planck15}:  ($\Omega_{\rm m}$,  $\Omega_{\Lambda}$,  $\Omega_{\rm b}$,  $h$, $n_s$,  $\sigma_8$) have the following values (0.308, 0.692, 0.0482, 0.678, 0.961, 0.829). The 80-2048 (160-1024) simulation follows the evolution of $2\times2048^3$ ($2\times1024^3$) cold dark matter plus baryon particles within a periodic box of linear comoving size of 80 (160) $h^{-1}$ Mpc from $z=99$ to $z=2.4$. 
In table \ref{tab:sims} we show the mass resolution and softening length for the two different runs. Here we briefly summarize the main physical ingredients of the hydrodynamics.

\begin{table}[h]
\begin{center}
    \begin{tabular}{|c|c|c|c|c|l}
    \hline
    Name     & Box size  & $m_{\rm CDM}$  &$m_{\rm baryon}$ & $l_{\rm soft}$\\ 
    & $[h^{-1} {\rm Mpc}]$ & $[ h^{-1}{\rm M}_{\odot}]$ & $[ h^{-1}{\rm M}_{\odot}]$ & $[h^{-1} {\rm kpc}]$ \\
    \hline \hline
    80-2048   & 80   &  $4.3\times10^6 $ & $7.8\times10^5 $    & 1.56       \\ \hline
    160-1024 & 160  & $2.8\times10^8$ & $5.1\times10^7$       & 6.25   \\ \hline
    \end{tabular}
    \caption{\label{tab:sims}Summary of the resolution parameters of the two different simulations.}
\end{center}
\end{table}

Star formation processes are not properly followed in the simulations: gas particles with temperature T $< 10^5$ K
and an overdensity  $ > 1000$ are converted to collisionless particles, resulting in a significant increase in computation speed at the expense of removing cold, dense gas from the model.  The simulations do not include metal line cooling. The photoionisation and photo-heating of the hydrogen and helium gas is calculated using the spatially uniform Haardt \& Madau (2012) ionising background model, where HI reionization happens at $z\sim 12$. Moreover, the gas is assumed to be optically thin and in ionisation equilibrium. 
We stress that 80-2048 and 160-1024 share the same thermal history, since for both of them the cooling routine and the UV background have been modified following the reference model of \cite{viel13}. 
These simulations are tailored to give converged properties for low density intergalactic medium statistics as  probed by the Ly$\alpha$ forest. For a detailed description of the simulations we refer the reader to the paper \cite{bolton2016}. 

We analyze snapshots at $z=2.4$ for both simulations, for which halos are identified using the Friend-of-Friends (FoF) \cite{FoF} algorithm with a linking parameter length of $b=0.2$. The redshift choice of $z=2.4$ is convenient for our purposes since it is within the redshift range for which the highest number of quasars is observed (hence most \lya forest data available, for example see \cite{slosar2011}) and for which 21cm observations are planned to be carried out (e.g. by SKA\footnote{https://www.skatelescope.org} \cite{braun15} or CHIME\footnote{http://www.mcgillcosmology.ca/chime}).

\begin{figure}
\begin{center}
\includegraphics[width=12cm]{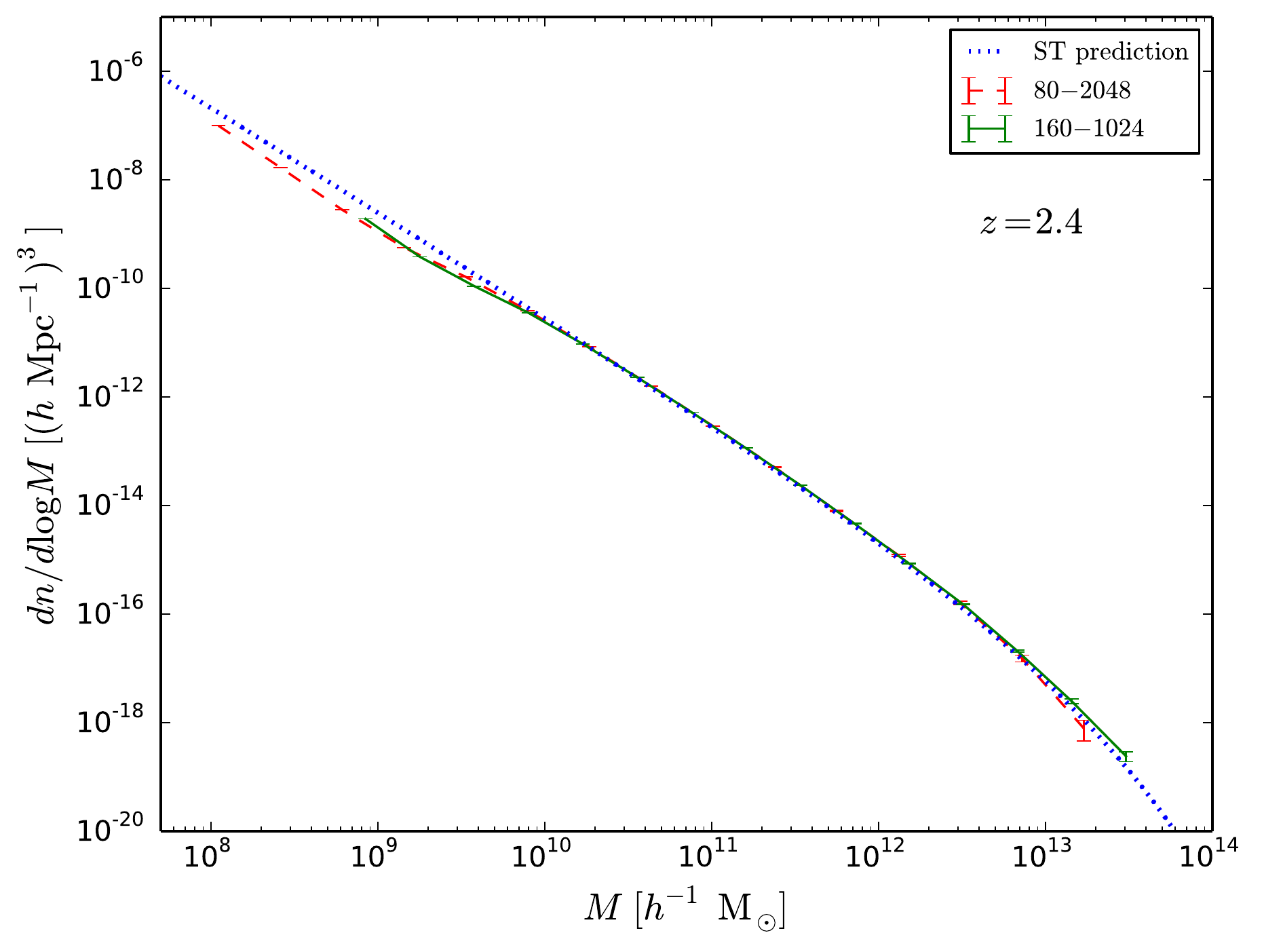}
\caption{Halo mass functions of the 80-2048 simulation (red dashed line) and of the 160-1024 simulation (green solid line) compared with the Sheth-Tormen prediction \cite{STmodel} (blue dotted line). The error bars are the uncertainties in each of the mass bins used for the computation assuming that the number of halos follows a Poissonian distribution.}
\label{fig:MF}
\end{center}
\end{figure}

We made use of two simulations in order to carry out convergence tests. In this paper we are interested in studying the shape and amplitude of the cross-signal on linear and mildly non-linear scales. Thus, the 160-1024 simulation would be the best suited for this analysis as its volume is eight times higher than the one of 80-2048. On the other hand, the mass and spatial resolution of the 80-2048 simulation is much higher than the one of the 160-1024, so we can investigate whether the ingredients we need to carry out our analysis are converged already in the 160-1024 simulation and therefore focus our analysis on that run.

As we discuss in detail in section \ref{sec:21cm_distribution} we model the 21cm signal using the halo information from the simulation catalogues and therefore we do not attempt to account for HI self-shielding and presence of molecular hydrogen for each gas particle in our simulations. In this sense, an N-body simulation would have been enough for our purposes, with the advantage to be less computational expensive to run and to reach larger scales more easily. On the other hand, a precise modelling of the Ly$\alpha$ forest requires high-resolution hydrodynamic simulations as we describe in detail in section \ref{sec:Lya}. In this paper we decided to use state-of-the-art hydrodynamic simulations that accurately model the properties of the Ly$\alpha$ forest at the expenses of neglecting the contribution of HI outside halos to the global 21cm signal (which was found negligible in \cite{villaescusa2014}) and avoiding modelling the intrinsic scatter in the $M_{\rm HI}(M,z)$ function, that represents the average HI mass that a dark matter halo of mass M hosts at redshift z. However, we notice that we are also  testing a method to model the Ly$\alpha$ signal in pure dark matter simulations (see appendix \ref{app}) and this will be  the subject of a future study.

As already stated, the halo information is crucial for this work since our 21cm maps are built from it. In Fig. \ref{fig:MF} we plot the halo mass function for the 80-2048 and 160-1024 simulations in red and green, respectively, while the dotted blue line shows the Sheth \& Tormen (ST) prediction \cite{STmodel} for the corresponding redshift  $z=2.4$. We find that the halo abundances from the simulations agree well with each other in the common mass range and also with the ST prediction. The higher mass resolution of the simulation 80-2048 allows us to sample the low mass end of the halo mass function while the larger box size of the 160-1024 simulation is better suited to explore the high-mass end. Thus, the combination of both simulations enables us to sample the halo mass function on a wide range of masses as shown in Fig. \ref{fig:MF} which is important for our modelling of the 21cm signal.

\section{Modelling the neutral hydrogen distribution}
\label{sec:21cm_distribution}

In this section we illustrate how we model the 21cm signal. First, in subsection \ref{sub:bagla} we describe the model for spatially distributing neutral hydrogen (HI), responsible for the 21cm emission. Next in \ref{sub:21Pk}, we explain the method we use to transform the spatial distribution of HI into 21cm maps, and we show a comparison of the clustering properties of the two simulations.

\subsection{The halo-based model}
\label{sub:bagla}

In the hydrodynamic SPH simulations the spatial distribution of gas is discretized into a finite number of gas particles with a given kernel and radius. The phase in which the gas is (ionized, neutral, forming molecules) can be found through radiative-transfer calculations. While this is the most robust way to model the spatial distribution of HI, it is also the most computationally expensive. Unfortunately, a radiative-transfer calculation will not output the correct HI distribution in our simulations. The reason for this is because the hydrodynamic simulations have been run using the so-called "quick-Ly$\alpha$" flag: the code follows the full hydrodynamic evolution of the gas until this reaches a given density and temperature threshold (see section \ref{sec:simulations} for further details); at that stage the code will transform the gas particle into a collisionless star. This technique allows to speed up calculations by avoiding modelling the gas in the inter-stellar medium (ISM), where most of the HI is located. This kind of simulations produce too many stars that make the gas reservoir unreliable.

In order to avoid the above problem we can model the spatial distribution of HI assuming that all HI is confined within dark matter halos. This is a reasonable assumption since the UV background will prevent the formation of large HI clouds unless they are self-shielded. Besides, this statement has been verified through hydrodynamic simulations in \cite{villaescusa2014}. Under these conditions it is possible to develop a HI halo model whose main ingredient is the function $M_{\rm HI}(M,z)$, that represents the average HI mass that a dark matter halo of mass $M$ hosts at redshift $z$. Therefore, instead of computing the hydrogen phase fractions of each gas particle, we can estimate the spatial distribution of HI by modelling the $M_{\rm HI}(M,z)$ function.

We emphasize that by modelling the $M_{\rm HI}(M,z)$ function we are implicitly neglecting the intrinsic scatter expected in it. However, the scatter will not affect the clustering properties of the HI since it only depends on the average value. On the other hand, if the scatter correlates with environment cross-correlations may be affected by it. We will investigate this issue in a subsequent paper. Here we assume that the scatter in the $M_{\rm HI}(M,z)$ does not correlate with environment and therefore we can avoid modelling it. To reinforce this point we notice that such halo-based models have been already extensively investigated at similar redshifts against other methods such as post-processing pseudo radiative-transfer calculations, and the results have been proven to be robust against the model used: e.g. \cite{villaescusa2014,villaescusa2015,carucci2015}.

We model the $M_{\rm HI}(M,z)$ as
\be
  M_{\rm HI}(M)=\begin{cases}
    f \, M^{\alpha}& \text{if $M_{\rm min}\leq M$}\\
    0 & \text{otherwise}.
  \end{cases}
 \label{eq:bagla}
\ee
where we set $\alpha=3/4$. This functional form arises from the result of high-resolution and zoom-in hydrodynamic simulations \cite{villa_neutrinos, villa_clusters} and it has been shown that it is capable of reproducing the abundance and clustering of the Damped Lyman-$\alpha$ systems (DLAs)  \cite{castorina_HI_bias}. $M_{\rm min}$ is a lower mass cut-off that considers that a minimum hydrogen density (clustered in a minimum halo potential well) is necessary to have hydrogen self-shielding and prevent the gas to be totally ionised. Here we assume that the mass parameter $M_{\rm min}$ corresponds to a dark matter halo with circular velocitiy $v_{\rm circ} = 25$ km ${\rm s}^{-1}$, calculated using the virial relation:
\be
M = 10^{10} {\rm M}_{\odot} \left( \frac{v_{\rm circ}}{60 \,{\rm km\,s}^{-1}} \right)^3 \left( \frac{1+z}{4} \right)^{-1.5} ~.
\ee

We choose the value of the free parameter $f$ by requiring that our model reproduces the measured value of $\Omega_{\rm HI}$:
\be
\Omega_{\rm HI}(z)=\frac{1}{\rho_c^0}\int_0^\infty n(M)M_{\rm HI}(M,z)dM~,
\ee
where $n(M)$ is the halo mass function and $ \rho_{\rm c}^0$ is the present day critical density of the universe. Since we have a particular realization of the halo field we impose the above condition in our simulations as
\be
f \sum_{i=0}^n M^{\alpha}\,\Theta(M_i-M_{\rm min}) = \Omega_{\rm HI} L^3 \rho_{\rm c}^0 \,,
\ee
where $L$ is the simulation box size, $\Theta(x)$ is the Heaviside step function and the index $i$ runs over all the dark matter halos of the simulation.

Since the value of $\Omega_{\rm HI}$ is poorly constrained by observations at redshift $z\simeq2.4$ (see \cite{ramirez2016} for a recent report) we set $\Omega_{\rm HI} = 10^{-3}$ in both simulations and notice that our conclusions are not affected by this choice as it only controls the amplitude of the 21cm power spectrum, not the HI clustering. We also notice that the value of $f$ derived for the two simulations agree well with each other, pointing out that the mass function of both simulations overlap and $M_{\rm min}$ is resolved in both.

While the clustering of HI on large-scales is fully determined by the function $M_{\rm HI}(M,z)$, on smaller scales it depends on the way the HI is distributed within halos. Here for simplicity we avoid modelling the density profile of the HI inside halos (see e.g. \cite{hamsa2016}), and place it all in the halo center. In the appendix \ref{appHI} we show that, for the scales we are interested in, this approximation does not bias our results.

\begin{figure}
\begin{center}
\includegraphics[width=\textwidth]{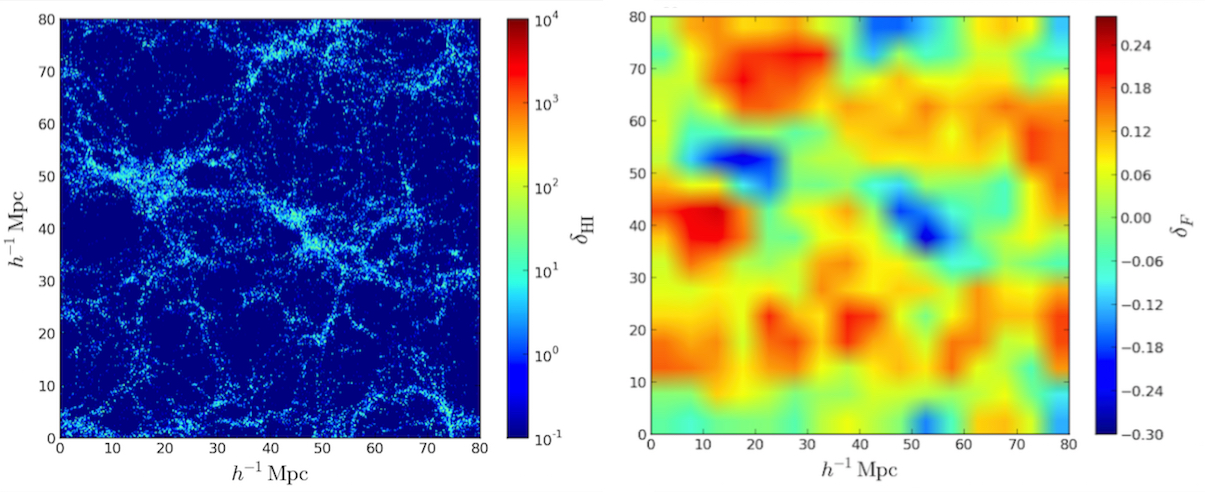}
\caption{Spatial distribution of the density contrast of HI in redshift-space (left) and of the density contrast of the \lya forest flux (right) at $z=2.4$. We show the whole 80-2048 simulation box, taking a slice of $10\,h^{-1}\,{\rm Mpc}$ width.}
\label{fig:snap}
\end{center}
\end{figure}

By looking at the HI spatial distribution in Fig. \ref{fig:snap}, it is evident how HI tracks the halo positions in the simulation box, thus revealing the web-like structure of the matter in the Universe. We notice that since we are not modelling the HI 1-halo term, the presence of fingers-of-God in Fig. \ref{fig:snap} is highly suppressed.

Since our model consists in assigning HI in a deterministic manner to dark matter halos of a given mass, we can compute the linear HI bias that we should recover in our simulations as
\be
b_{\rm HI}(z)=\cfrac{\int_0^\infty b(M,z)n(M,z)M_{\rm HI}(M,z)dM}{\int_0^\infty n(M,z)M_{\rm HI}(M,z)dM}~,
\ee
where $b(M,z)$ is the halo bias. By using our HI model together with the halo mass function and halo bias from Sheth \& Tormen \cite{ST} and Sheth, Tormen \& Mo \cite{SMT}, we get a value for the HI bias equal to $b_{\rm HI}(z=2.4)=1.45$, in disagreement with the constraints on the bias of the DLAs \cite{font-ribera2012}; however, the purpose of this work is to be able to retrieve the input HI model values when fitting the results of the simulations. We also emphasize that a higher value of the HI bias will turn out to make the HI bias more scale-dependent \cite{villaescusa2014,Santos_2015}, and therefore shifting to large scales the onset of non-linearities, a situation we want to avoid given the relatively small volume of our simulations.

\subsection{The 21cm signal}
\label{sub:21Pk}

Radio telescopes are sensitive to the 21cm radiation emitted by HI. Therefore, the quantity directly measured in these observations is not the HI power spectrum, but the 21cm power spectrum, which is nothing but the power spectrum of the spatial distribution of neutral hydrogen in redshift-space, times an overall normalization factor. In this subsection we go through the steps for converting the HI distribution to the observed signal, the so called brightness temperature contrast: the radiation temperature against the CMB one at the redshift of observation. 

After obtaining the HI distribution in comoving real-space as described in the previous subsection \ref{sub:bagla}, we first move to redshift-space by displacing the position $\vec{x}$ of the particles (or of the halos) to $\vec{s}$ as
\be
\vec{s}=\vec{x}+\frac{1+z}{H(z)} \vec{v}_{\rm los}(\vec{x})\,,
\ee
with $z$ being the redshift of observation, $\vec{v}_{\rm los}$ the line of sight component of the peculiar velocity and $H(z)$ the Hubble parameter. Then, we compute the brightness temperature fluctuations using \cite{mao2012}:
\begin{equation}
\delta T_b(\vec{s})=\overline{\delta T_b}(z)\left[\frac{\rho_{\rm HI}(\vec{s})}{\bar{\rho}_{\rm HI}}\right],
\end{equation}
where $\rho_{\rm HI}(\vec{s})$ is the HI density in the redshift-space, $\bar{\rho}_{\rm HI}$ the HI mean density and 
\begin{equation}
\overline{\delta T_b}(z)=\Omega_{\rm HI}\left( \frac{23.88 h^2}{0.02 } \frac{100}{76}\right)\sqrt{\frac{0.15}{\Omega_{\rm m}h^2}\frac{(1+z)}{10}}~{\rm mK}.
\label{eq:deltaTb}
\end{equation}
Finally, the 21cm power spectrum is defined as $P_{\rm 21cm}(k)=\langle \delta T_b(\vec{k}) \delta T_b^*(\vec{k})  \rangle$. Notice that the amplitude of the signal depends on the total amount of cosmic neutral hydrogen, expressed by the density parameter $\Omega_{\rm HI}$ in Eq. \ref{eq:deltaTb}.

As we discussed before, in our formalism we do not model the HI density profile. In the appendix \ref{appHI} we show the differences arising in the 21cm power spectrum by the model used to describe the HI density profile.

\begin{figure}
\begin{center}
\includegraphics[width=12cm]{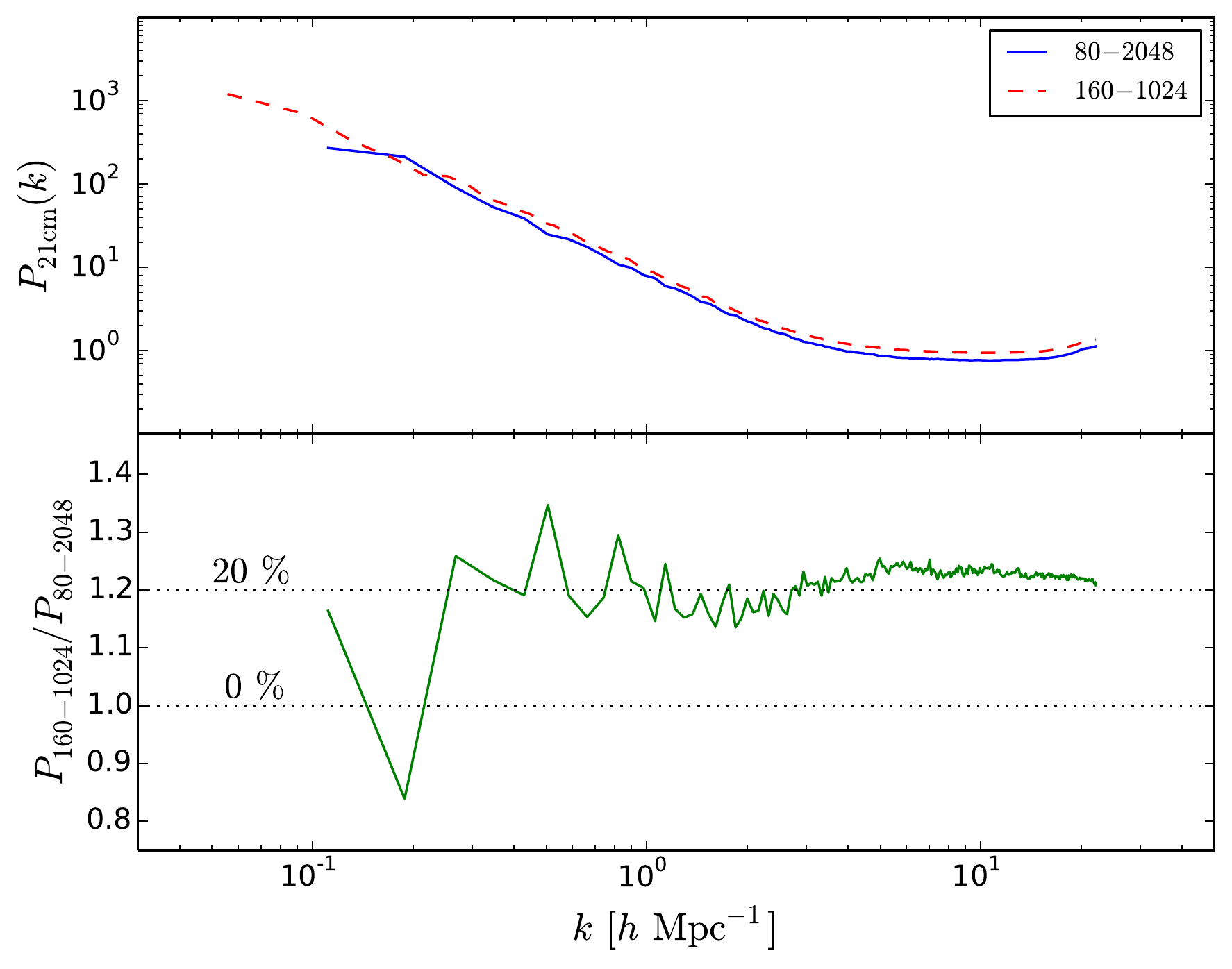}
\caption{The upper panel displays the 21cm power spectrum from the 80-2048 (solid blue) and 160-1048 (dashed red) simulations at $z=2.4$ while the lower panel shows their ratio.}
\label{fig:21cmPk}
\end{center}
\end{figure}

We do a direct comparison between the two simulation 21cm power spectra in Fig. \ref{fig:21cmPk}, upper panel, with the residuals $P^{160-1024}_{\rm 21cm}(k)$ over $P^{80-2048}_{\rm 21cm}(k)$ in green in the lower panel. We find that the shape of both power spectra agree on all scales, with a $\sim20\%$ offset in amplitude among the two: HI is more clustered in the big box simulation. The reason for this amplitude difference is due to the fact that the larger box is able to capture the large-scale modes, which are not present in the low volume box. The presence of those modes enhances the clustering on lower scales, producing the effect we find here.

\section{Modelling the Ly$\alpha$ forest}
\label{sec:Lya}

\begin{figure}
\begin{center}
\includegraphics[width=12cm]{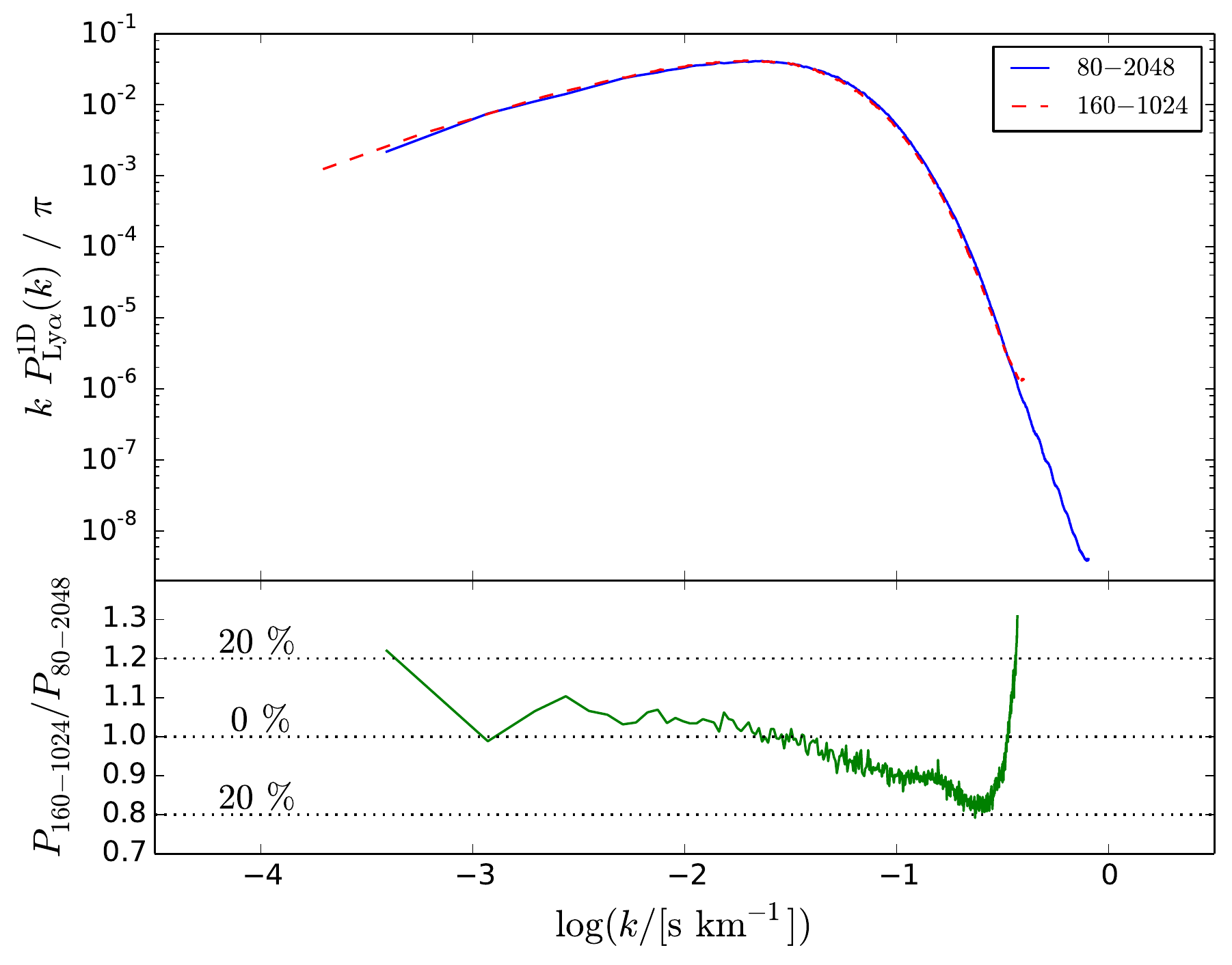}
\caption{{\it Upper panel:} The dimensionless 1D Ly$\alpha$ flux power spectrum for the 80-2048 (blue solid) and the 160-1024 (red dashed) simulations. {\it Lower panel:} Ratio of the two power spectra; we highlight the $20\,\%$ difference region.}
\label{fig:1dlyaPk}
\end{center}
\end{figure}

As we discussed above, we use state-of-the-art hydrodynamic simulations to model the gas properties in low-density environments. We extract mock Ly$\alpha$ absorption spectra skewers from the simulations. We have three different catalogues, containing 2500, 1600 and 900 spectra that are obtained from regular grids of $50^2$, $40^2$ and $30^2$ points along the $x$, $y$ and $z$ directions, respectively (in appendix \ref{app} we discuss how regularizing the spectra position on a grid does not bias our results). Each spectrum contains 2048 pixels, evenly distributed along the simulation box length.

The gas density, weighted temperature, the neutral fraction and gas peculiar velocities are extracted following the SPH interpolation scheme described in \cite{theuns1998}; for the Ly$\alpha$ optical depth $\tau$ along each line of sight we make use of the Voigt profile approximation, as in \cite{tepper2006}. Once $\tau$ is determined in every pixel, we define the absorption flux as $F = e^{- \tau}$: this is the field we consider. For computing the power spectrum $P_{{\rm Ly}\alpha}(k)$ we compute the flux contrast as 
\be
\delta_F(\vec{x}) = \f{F(\vec{x}) - \langle F \rangle}{\langle F \rangle}
\ee
where $\langle F \rangle$ is the flux mean (see appendix \ref{app} for more details). For a more detailed description of the 160-1024 absorption spectra and comparison with observational data in terms of \lya forest statistics we refer the reader to \cite{bolton2016}.

\subsection{The Ly$\alpha$ forest auto-power spectrum}
\label{sec:lyaPk}

\begin{figure}
\begin{center}
\includegraphics[width=12cm]{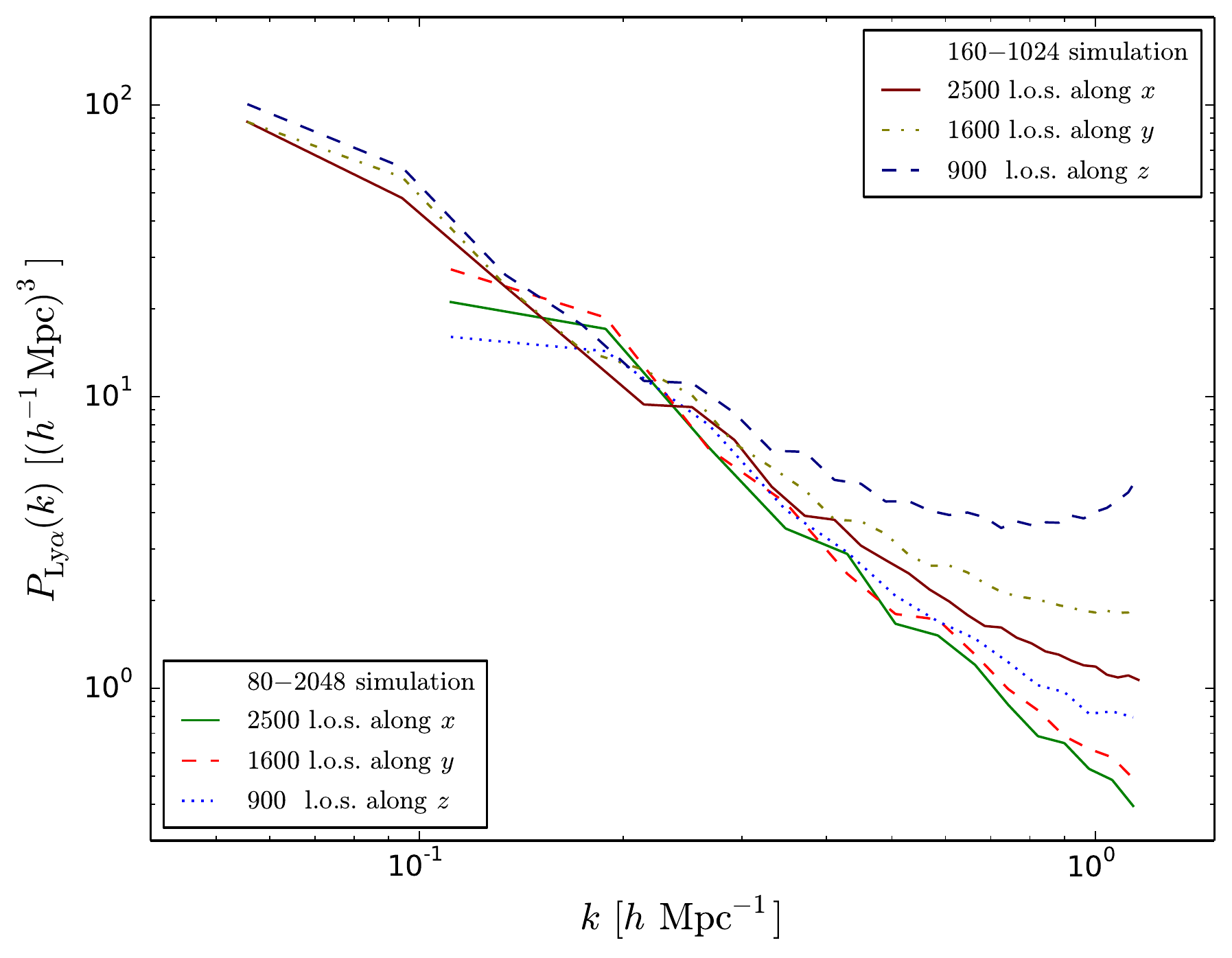}
\caption{The 3D Ly$\alpha$ flux power spectrum computed for the 3 different catalogues (with spectra along different directions) for each simulation.}
\label{fig:3dlyaPk}
\end{center}
\end{figure}

We first consider the one-dimensional Ly$\alpha$ power spectrum $P^{1{\rm D}}_{{\rm Ly}\alpha}(k)$. For each spectrum we compute the flux contrast $\delta_F$ in every pixel along the line of sight and the flux power spectrum. In Fig. \ref{fig:1dlyaPk} we show in the upper panel the two dimensionless power spectra, $k\,P^{1{\rm D}}_{{\rm Ly}\alpha}(k) / \pi$; with solid blue line for the 80-2048 simulation and in dashed red for the 160-1024. The shown power spectra have been computed by averaging the measured power spectrum from each individual skewer from the catalogue (5000 spectra in total: 2500 l.o.s. along $x$ plus 1600 along $y$ plus 900 along $z$). We find that our results are basically converged against resolution, as can be seen from the bottom panel, where we show their ratio, which stays within the $20\%$ difference region highlighted within dotted lines.

Next we compute the 3D Ly$\alpha$ power spectrum $P_{{\rm Ly}\alpha}(k)$, shown in Fig. \ref{fig:3dlyaPk}. There is good overlap among the six power spectra (giving that these catalogues have not been normalised, see appendix \ref{app}), especially on large scales, with the 80-2048 spectra slightly flattening at  $k \simeq 0.2 h \,{\rm  Mpc^{-1}}$. At smaller scales the 160-1024 spectra display an increase of power compared to the 80-2048 spectra because of the lack of small scale information, and the same effect is visible also within each simulation $P_{{\rm Ly}\alpha}(k)$: as the number of spectra per catalogue decreases (i.e. employing a smaller number of skewers in the box) we see an increase of power due to a sub-sampling of the field. 

Given the good agreement on both the 1D and 3D Ly$\alpha$ power spectra among the two simulations (on large-scales for the 3D power spectrum and on all scales in the 1D case) we conclude that the spatial distribution of gas in the Ly$\alpha$ forest is converged against resolution in the simulation 160-1048. We will thus focus our analysis on that simulation, since its larger box size allow us to explore larger scales than the 80-2048 simulation.

\section{Results}
\label{sec:results}

In this section, we start by computing the cross-correlation of the 21cm and Ly$\alpha$ forest signals in \ref{sub:X}. Next, in subsection \ref{sub:error} we check whether the detection of the cross-signal is possible in terms of realistic uncertainties on the measurements, whereas in \ref{sub:fit} we infer the respective bias parameters by analysing the power spectra within a linear theory framework.

In the analysis of this section, we use the 160-1024 simulation since we are interested in investigating the shape and amplitude of the cross-power spectrum on linear and mildly non-linear scales, which are better probed by that simulation.

\subsection{The 21cm-Ly$\alpha$ cross-correlation}
\label{sub:X}

\begin{figure}
\begin{center}
\includegraphics[width=12cm]{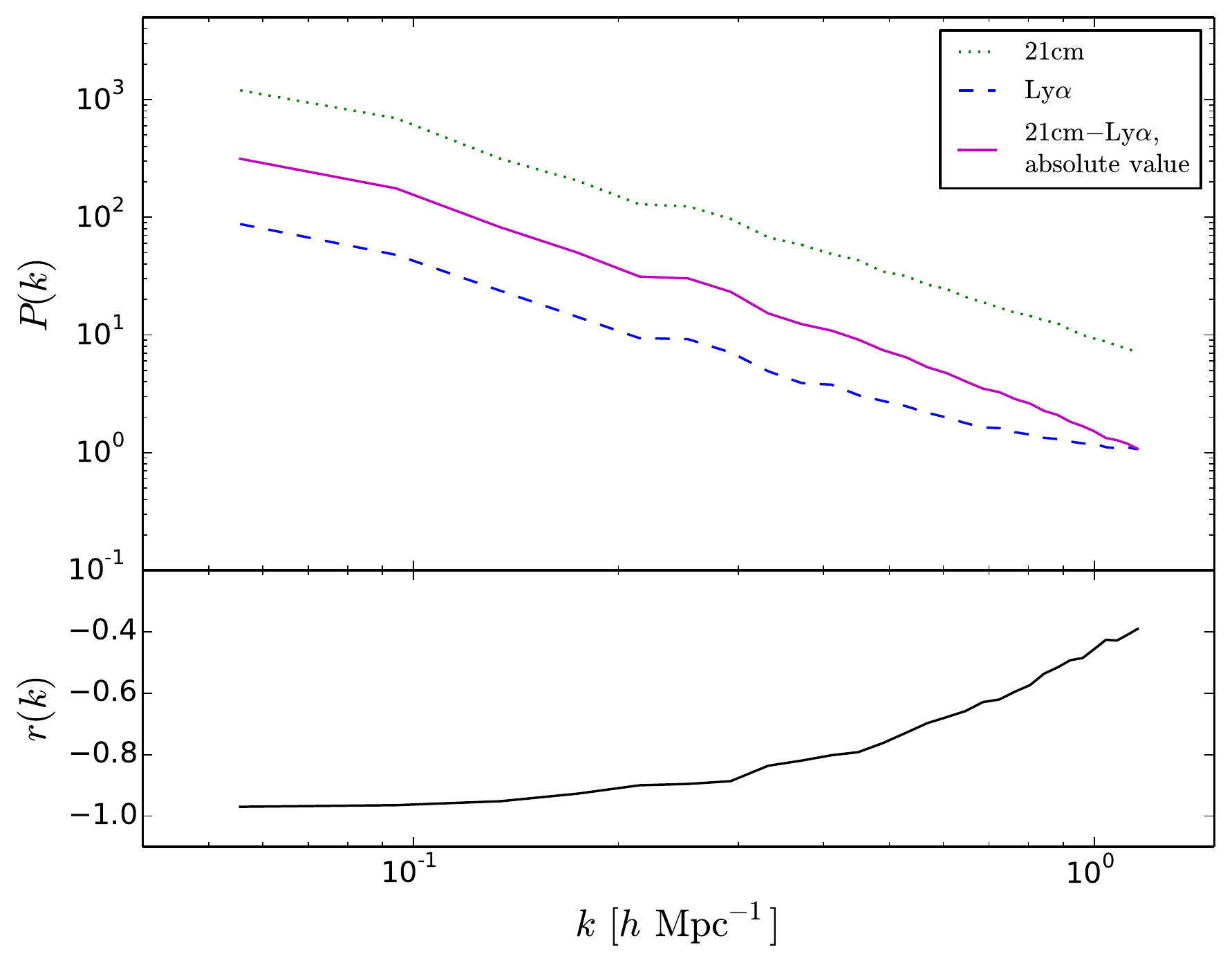}
\caption{The upper panel displays the power spectrum of the 21cm signal (dotted green), of the Ly$\alpha$ forest flux (dashed blue) and of their cross-correlation in absolute value (solid magenta). The bottom panel shows the cross-correlation coefficient, $r(k)$, among the two fields.}
\label{fig:cross}
\end{center}
\end{figure}

In Fig. \ref{fig:cross} we show with a dotted green line the 21cm power spectrum $P_{21{\rm cm}}(k)$, in dashed blue the 3D Ly$\alpha$ flux power spectrum $P_{{\rm Ly}\alpha}(k)$ and in solid magenta the absolute value of the 21cm-Ly$\alpha$ cross-power spectrum $P_{\rm X}(k)$\footnote{We emphasize that we are neglecting instrumental effects such as beam size or system noise as well as the impact of galactic and extragalactic foregrounds.}.  

We have plotted the absolute value of the cross-power spectrum because $P_{\rm X}(k)$ is negative, i.e. the fields are anti-correlated. Qualitatively, this results agrees with the picture of having the HI responsible for the 21cm radiation in dense environments as galaxies in halos, whereas the Ly$\alpha$ forest arises mainly from low-density, highly ionized, gas clouds in the intergalactic medium mostly residing in the filaments of the cosmic web. We will thus expect underdensities in the Ly$\alpha$ flux in places where 21cm overdensities are located (e.g. halos); this explains why the two fields are anti-correlated. We notice that this result agrees with recent observations by Mukae et al. \cite{mukae2016}, where they analyze HI-rich galaxies at $z \sim 2-3$ together with Ly$\alpha$ forest spectra finding anti-correlation between the two fields.

In the bottom panel of Fig. \ref{fig:cross} we plot the cross-correlation coefficient, defined as
\be
r(k)=\frac{P_{\rm X}(k)}{\sqrt{P_{\rm 21cm}(k)P_{{\rm Ly}\alpha}(k)}}~.
\ee
As expected from the arguments given above, we find the sign of the cross-correlation to be negative on all the scales, indicating that the fields are anti-correlated. On the largest scales probed by our simulation the value of the cross-correlation is close to $-1$, while for scales $k\ge0.2~h{\rm Mpc}^{-1}$ the value of the cross-correlation increases. We can naively associate this scale to non-linearities, as linear theory predicts a scale-independent cross-correlation coefficient. Thus, since in this paper we are interested in extract cosmological information from linear scales\footnote{We notice that perturbation theory is a powerful method to model mildly non-linear scales.}, we limit our analysis to modes with wavenumbers $k<0.2~h{\rm Mpc}^{-1}$.

In a nutshell: we are looking at signals coming from different regions in the sky. By considering together 21cm radiation and Ly$\alpha$ flux, we are looking at HI both in emission and in absorption, thus probing its cosmological amount and its spatial distribution at the same time. By visually inspecting the two fields in Fig. \ref{fig:snap}, we indeed notice how HI-poor regions (left panel, dark blue) correspond to high \lya forest transmitted flux regions (right panel, yellow and red).

To quantify the improvement gained by looking for HI information in cross-correlation, in next section we make predictions on the accuracy with which one should be able to make such measurements, taking care of the system noise only, ignoring effects arising from calibration, astrophysical foregrounds, variations in the ionosphere, radio-frequency interference, etc, which are beyond the scope of this paper.

\subsection{Error estimation and forecasts for SKA}
\label{sub:error}

In this section we describe the way we have computed the Gaussian errors for the measurements of the different power spectra of the previous section, and use the formalism to 

In this section we estimate the signal-to-noise ratio of a measurement of the auto- and cross-power spectrum of the Ly$\alpha$ forest and the 21cm fields performed on the volume where that overlaps observations from SKA1-MID \cite{braun15} and a BOSS-like survey \cite{dawson13} for the Ly$\alpha$ flux field. We notice that in this work we just focus on the cosmological signal, i.e. we neglect contributions from residual foreground contamination.

The accuracy with which one can measure the power spectrum $P_a(\vec{k})$, where $a$ stands for 21cm, Ly$\alpha$ or 21cm-Ly$\alpha$, is generally quantified by the signal to noise ratio $S/N$:
\be
\left(\f{S}{N}\right)^2_a(\vec{k}) = N_k \f{P^2_a(\vec{k})}{\sigma^2[P_a(\vec{k})]},
\label{eq:SN1}
\ee
where $N_k$ is the number of modes in each given bin centred at $(k,{\rm cos}(\theta))$, where $\theta$ is the angle between $\vec{k}$ and the line of sight. Here for concreteness and clearness we focus on the errors of the monopoles, but notice that the extensions to quadrupoles and monopoles-quadrupoles covariances is straightforward. For monopoles we average the amplitude of the power spectrum of modes with $k\in[k,k+dk]$, thus, the S/N ratio becomes:
\be
\left(\f{S}{N}\right)^2_a =  \f{2 \pi k^2 \de k V_{\rm survey}}{(2 \pi)^3} \int_0^{\f{\pi}{2}}\f{P^2_a(k,\theta){\rm sin}(\theta) \de \theta}{\sigma^2[P_a(k,\theta)] },
\label{eq:SN2}
\ee
where $V_{\rm survey}$ represents the survey volume. In our case we take it to be the volume of the region where both Ly$\alpha$ observations and the 21cm survey overlap. In particular, for a 21cm detection experiment we can write $V_{\rm survey} = D^2 \Delta D (\lambda^2 / A)$, where $D$ is the comoving distance to the redshift of observation, $\Delta D$ the comoving distance associated with the bandwidth of the instrument\footnote{We choose this bandwidth to match the common volume of both surveys.} $B = 32\,{\rm MHz}$, $\lambda$ the wavelength of observation (corresponding 21cm line at the redshift of detection) and $A=15$ m the collecting area of a single antenna.

We can cast the error on the 21cm detection via interferometric observations as\footnote{We notice that we are neglecting the contribution to the error from shot-noise. In \cite{castorina_HI_bias} it was shown that for the relevant scales we are interested here this term is subdominant.}:
\be
\sigma^2[P_{21{\rm cm}}(k,\theta)] = \left[ P_{21{\rm cm}}(k,\theta) + \f{T^2_{\rm sys}}{2 B t_0} \f{D^2 \Delta D}{n(k_{\perp})} \left(\f{\lambda^2}{A_e}\right)^2 \right]^2,
\label{eq:sigma21}
\ee
where $T_{\rm sys}$ is the system temperature of the radio telescope which is the sum of the temperature of the sky at this redshift $T_{\rm sky}  \simeq 60 (300{\rm MHz}/\nu_{\rm HI}(z))^{2.55}$, with $\nu_{\rm HI}(z)=1420/(1+z)$ MHz, the temperature receiver $T_{\rm rcvr} = 0.1 T_{\rm sky} + T_{\rm inst}$, with $T_{\rm inst} = 28{\rm K}$ the SKA1-MID instrument temperature and. $n(k_{\perp})$ is the number density of the interferometer baselines sensitive to the transverse mode $k_{\perp}$ which depends on the spatial distribution of the antennae that we calculate using the SKA1-MID baseline density distribution in \cite{villaescusa2015}. $A_e$ is the effective collective area of a single antennae: for SKA1-MID $A_e=140\,{\rm m}^2$. The parameter that we can tune is the total observation time $t_0$ that we conservatively set to 100 hours.

We express the error on the 3D Ly$\alpha$ flux power spectrum as a combination of the noise term with an aliasing term, due to the sparse sampling of the Ly$\alpha$ field made by the discrete lines of sight \cite{font-ribera2014}:
\be
\sigma^2[P_{{\rm Ly}\alpha}(k,\theta)] = \left[ P_{{\rm Ly}\alpha}(k,\theta) + P^{1{\rm D}}_{{\rm Ly}\alpha}(k\,{\rm cos}\theta) n_{\rm eff}^{-1} \right]^2.
\ee
The aliasing term contributes in the line of sight direction with the 1D flux power spectrum $P^{1{\rm D}}_{{\rm Ly}\alpha}(k\,{\rm cos}\theta)$ multiplied by a noise-weighted density of lines of sight per unit area $n_{\rm eff}$. From \cite{slosar2011} we know that for a redshift bin to which $z=2.4$ belongs the lines of sight density for BOSS is $n_{\rm eff} \simeq 15\, {\rm deg}^{-2}$. We choose to use both this value and its double ($30 \,{\rm deg}^{-2}$) to estimate errors for a BOSS-like and a next generation BOSS-like survey like DESI \footnote{http://desi.lbl.gov/}. Using the conversion table in \cite{McQuinn&White2011}, we obtained the values $n_{\rm eff} = 0.003\,{\rm and}\,0.006\,(h^{-1}\,{\rm Mpc})^{-2}$, respectively.

Finally, the error on the measurement of the cross-correlation power spectrum can be written as: 
\be
\sigma^2[P_{\rm X}(k,\theta)] = \f{1}{2} \left( P^2_{\rm X}(k,\theta)+ \sigma[P_{21{\rm cm}}(k,\theta)]  \sigma[P_{{\rm Ly}\alpha}(k,\theta)] \right).
\label{eq:sigmaX}
\ee

On small scales, the error budget of the 21cm power spectrum is dominated by the system noise and, by looking at the formula \ref{eq:sigma21}, we know it scales with the observing time as $1/t_0$, whereas for the cross-correlation power spectrum (see equation \ref{eq:sigmaX}) goes as $1/\sqrt{t_0}$, so if we choose a more  optimistic survey observing time of $t_0=1000$ hours instead of 100 hours, $\sigma[P_{21{\rm cm}}(k)]$ would improve by a factor 10 and $\sigma[P_{\rm X}(k)]$ would be $\sim$ 3 times smaller. Anyways, at the scales we are looking at ($k<0.2 h \,{\rm  Mpc^{-1}}$) this is no longer valid: observing for longer does not beat cosmic variance. Hence, we keep $t_0=100$ hours for our analysis and we do not explore other observing time possibilities.

\begin{figure}
	\begin{center}
		\includegraphics[width=12cm]{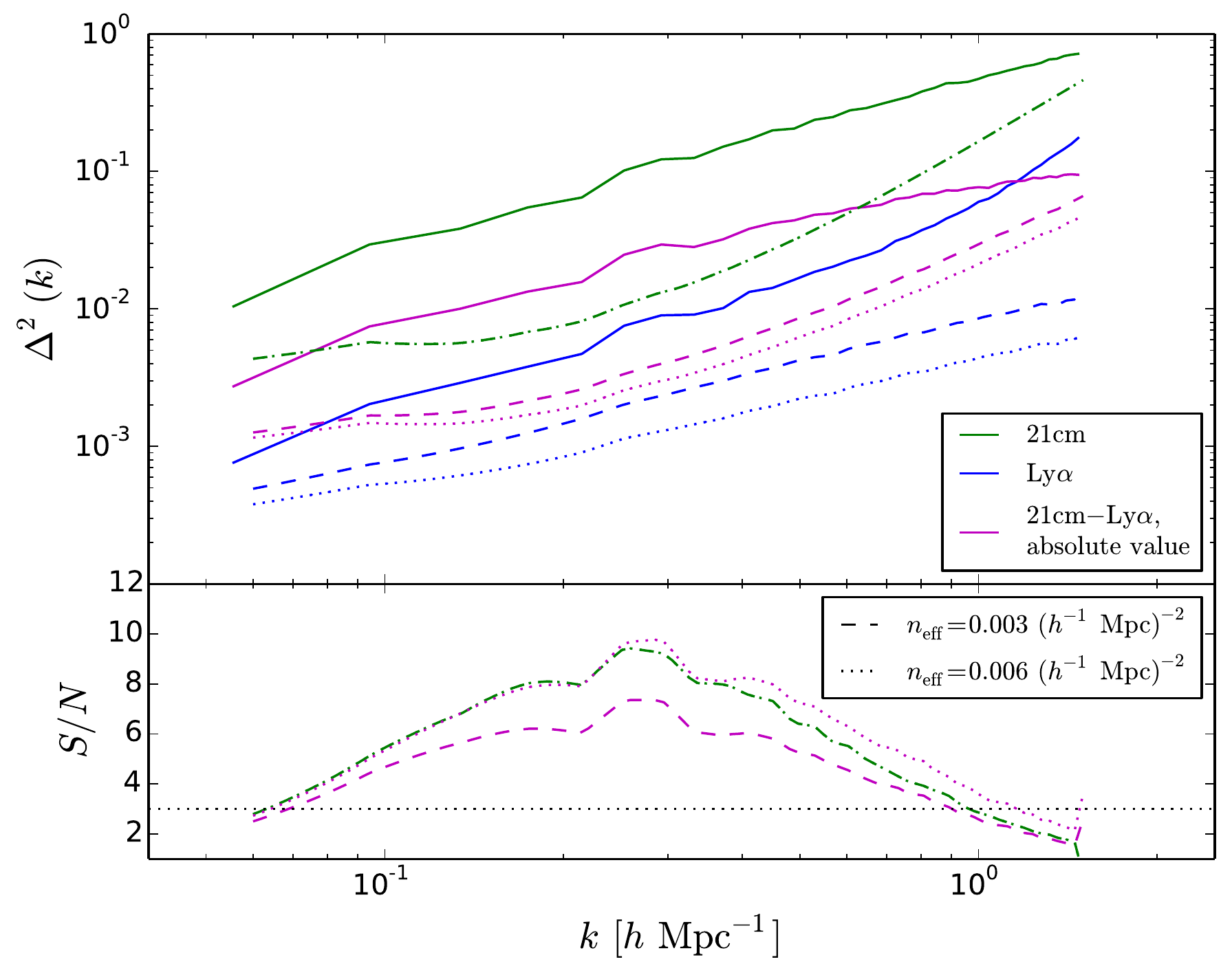}
		\caption{{\it Upper panel:} The dimensionless power spectrum $\Delta^2(k)=k^3 P(k) / 2 \pi^2$ in solid lines for the 21cm radiation (green), the Ly$\alpha$ forest flux (blue) and for their cross-correlation in absolute value (magenta). The non-solid lines are the estimated errors on the power spectra for a SKA1-MID (21cm intensity mapping) and BOSS like (Ly$\alpha$ flux) surveys, the latter with an effective density of lines of sight $n_{\rm eff} = 0.003 (h^{-1} \,{\rm  Mpc})^{-2}$ in dashed lines or $n_{\rm eff} = 0.006 (h^{-1} \,{\rm  Mpc})^{-2}$ in dotted lines. See section \ref{sub:error} for details. {\it Lower panel:} Signal-to-noise ratio defined as the ratio between the power spectrum and its error. The horizontal dotted line marks $S/N = 3$.}
		\label{fig:SNR}
	\end{center}
\end{figure}

We use the above formalism to compute the errors on the multipoles of the different power spectra, and their covariances, for the measurements we carried out in the previous section. We notice that since the scales we are interested when performing the fit are almost linear ($k<0.2~h{\rm Mpc}^{-1}$), the error budget is dominated by cosmic variance. Thus, in order to speed calculations up we have neglected the contribution of system noise to the 21cm power spectrum when computing the errors on the power spectra of the previous section. This is the reason of the low amplitude of the errors on the 21cm power spectra in Fig. \ref{fig:fits}, but notice that our results do not change if we include them. 

However, we account for the contribution of system noise when estimating the S/N ratio for our forecasts in this section. We summarize the errors estimation in Fig. \ref{fig:SNR}, where we plot their magnitude together with the amplitude of the different auto- and cross-power spectra.

In the upper panel of Fig. \ref{fig:SNR} we plot the dimensionless power spectrum $\Delta^2_a(k) = k^3 P_a(k) / (2 \pi^2)$ with solid lines: green for the 21cm radiation, blue for the Ly$\alpha$ flux and magenta for the absolute value of their cross-signal, together with the errors $\sigma[P_a(k)]$ in dashed lines using $n_{\rm eff} = 0.003\,(h^{-1}\,{\rm Mpc})^{-2}$ and dotted lines for $n_{\rm eff} = 0.006\,(h^{-1}\,{\rm Mpc})^{-2}$. Clearly, the error on the 21cm power spectrum does not depend on the value of  $n_{\rm eff}$, and the improvement on the $P_{21{\rm cm-Ly}\alpha}(k)$ is evident but smaller than that on $P_{{\rm Ly}\alpha}(k)$. Nonetheless, we can notice that the aliasing term in $\sigma^2[P_{{\rm Ly}\alpha}(k)]$ (and consequently in $\sigma^2[P_{\rm X}(k)]$) dominates the variance budget: therefore the way to do better would be to conduct more sensitive Ly$\alpha$ spectra surveys, i.e. by increasing the surveyed volume and/or increasing the quasars number density.

In the bottom panel of Fig. \ref{fig:SNR} we plot the signal to noise ratio $S/N$ for $P_{21{\rm cm}}(k)$ and $P_{21{\rm cm-Ly}\alpha}(k)$. We find that S/N ratio for both power spectra peak around $k\simeq0.25~h{\rm Mpc}^{-1}$. On larger scales cosmic variance dominates the total error, while on smaller scales the noise from the instrument starts dominating. 

We thus conclude that the cross-power spectrum of the Ly$\alpha$ forest from a BOSS like survey with 21cm intensity mapping interferometry observations in the post-reionization era from the SKA1-MID instrument can be detected with a large S/N ratio on scales $k\in[6\times10^{-3}-1]~h{\rm Mpc}^{-1}$ with a very conservative total observing time of 100 hours. We notice that these numbers should be regarded as lower limits, since our simplistic model for $\rho_{\rm HI}(r|M,z)$ will underestimate the clustering on the 1-halo term.

The amplitude and shape of the cross-power spectrum on large-scales only depends on the value of the bias and redshift-space distortion parameters of both fields. In the following section we check whether by analysing our simulated data we are able to retrieve the bias parameters and how well linear theory is able to describe the simulated fields.

\subsection{Linear theory comparison and bias parameters estimation}
\label{sub:fit}

In this section we introduce the theoretical model for the auto- and cross-power spectra and use it to blindly estimate the value of the bias parameters of both the Ly$\alpha$ forest and the 21cm fields. We also investigate how much information we gain by using the cross-power spectrum together with the auto-power spectrum measurements.

At linear order, the amplitude and shape of the 21cm, Ly$\alpha$ flux and 21cm-Ly$\alpha$ power spectra in redshift-space can be expressed as:
\bear
\label{eq:21cm_lin}
P_{\rm 21cm}(k,\mu) &=& A^2 \Omega_{\rm HI}^2 b_{\rm HI}^2 \left( 1+\beta_{\rm HI} \mu^2 \right)^2 P_{\rm m}(k) ,\\
\label{eq:Lya_lin}
P_{{\rm Ly}\alpha}(k,\mu) &=& b_F^2 \left( 1+\beta_F \mu^2 \right)^2 P_{\rm m}(k) ,\\
P_{\rm X}(k,\mu) &=& A \Omega_{\rm HI} b_{\rm HI} \left( 1+\beta_{\rm HI} \mu^2 \right) b_F \left( 1+\beta_F \mu^2 \right) P_{\rm m}(k) ,
\label{eq:cross_lin}
\ear
where the normalisation factor $A$ is computed from Eq. \ref{eq:deltaTb} defining $\overline{\delta T}_b=A \Omega_{\rm HI}$, with $\Omega_{\rm HI}=10^{-3}$ in our modelling, $b_{\rm HI}$ and $\beta_{\rm HI}$ are the linear bias and the redshift-space distortion parameter for the 21cm field, $b_F$ and $\beta_F$ those for the Ly$\alpha$ forest, $P_{\rm m}(k)$ is the linear matter power spectrum and $\mu$ is the cosine of the angle between the Fourier mode vector {\bf k} and the line of sight.

From the numerically computed $P_{\rm 21cm}(\vec{k})$, $P_{{\rm Ly}\alpha}(\vec{k})$ and $P_{\rm X}(\vec{k})$, we determine the values of the four bias parameters ($b_F$, $\beta_F$, $b_{\rm HI}$, $\beta_{\rm HI}$) using two different methods.

{\bf Method 1: auto-power spectrum multipoles.} This method consists in determining the value of the two bias parameters of each observable through fitting the monopole and quadrupole of each respective field in redshift-space. The power spectrum multipoles can be computed from the 2D power spectrum as:
\be
\label{eq:legendre}
P_l(k) = \f{2l+1}{2} \int_{-1}^{1} P(k,\mu) L_l(\mu) d\mu \,,
\ee
where $L_l(\mu)$ are Legendre polynomials. Thus, we can express the monopole and quadrupole as:
\bear
\label{eq:multipoles}
P_{a,0}(k)&=&b_a^2 \left( 1+\f{2}{3}\beta_a+\f{1}{5}\beta_a^2  \right) P_{\rm m}(k) ,\\
P_{a,2}(k) &=&b_a^2 \beta_a \left( \f{4}{3} +\f{4}{7}\beta_a \right) P_{\rm m}(k)  ,
\label{eq:multipoles2}
\ear
where $a$ stands either for 21cm or for Ly$\alpha$ (and here we omitted the normalisation factors for the 21cm case, see above Eq. \ref{eq:21cm_lin}). If we were interested in the value of $\beta_a$ only, we could extract it from the ratio between quadrupole and monopole, which depends only on this parameter; to constrain also $b_a$, we need to assume a cosmological model through $P_{\mathrm m}(k)$. The 21cm redshift-space distortion parameters contain information on the growth rate, $f$, since $f=\beta_{\rm HI}b_{\rm HI}$; the latter relation does not hold for the Ly$\alpha$ case, due to its intrinsically non-linear correspondence to the underlying matter density field \cite{slosar2011} and this is the reason why adding information from the cross-power spectrum cannot directly improve the measurement of the linear growth rate, but it tightens the constraints on the ($b_{\rm HI}$, $\beta_{\rm HI}$) parameters as we later show.

We do the best fit to Eqs. \ref{eq:multipoles}-\ref{eq:multipoles2} via a Monte Carlo Markov Chain (MCMC) on the two free parameters ($b_a$ and $\beta_a$) making use of the \emph{emcee} package \cite{emcee}. Errors on the monopoles and quadrupoles of the auto-power spectra are computed assuming the modes follow a Gaussian distribution (see appendix \ref{app_errors}), which is a good description on large, linear scales, and just taking into account the contribution from cosmic variance. We consider a cosmological volume equal to the one probed by our simulations and using only modes with $k<0.2~h{\rm Mpc}^{-1}$. We also account for the correlation between the monopoles and quadrupoles in the covariance matrix. We notice that when performing the fit we are neglecting the correlation between multipoles of the two fields through the cross-power spectrum. The best-fit values together with their $1\sigma$ errors are shown in the upper row of table \ref{tab:fit}. 

The recovered values for the HI, $b_{\rm HI}=1.520^{+0.058}_{-0.060}$ and $\beta_{\rm HI}=0.720^{+0.14}_{-0.13}$ are in agreement with the input ones, $b_{\rm HI}=1.45$, $\beta_{\rm HI}=0.67$, at $\sim1\sigma$. Notice that we have not inputed any model for the Ly$\alpha$ forest, whose properties are directly extracted from the output of the simulations. We find $\beta_F=1.480^{+0.21}_{-0.20}$, $b_F=-0.144^{+0.007}_{-0.007}$, while from observations it has been measured $\beta_{\rm F}=1.39 \pm 0.1$ and $b_{\rm F} (1+\beta_{\rm F})=-0.374 \pm 0.007$ \cite{Blomqvist2015}, thus, in perfect agreement. We notice that we obtain a good normalized $\chi^2$ of $13.6/12$, showing the model is a good description of the data.

\begin{figure}
\begin{center}
\includegraphics[width=12cm]{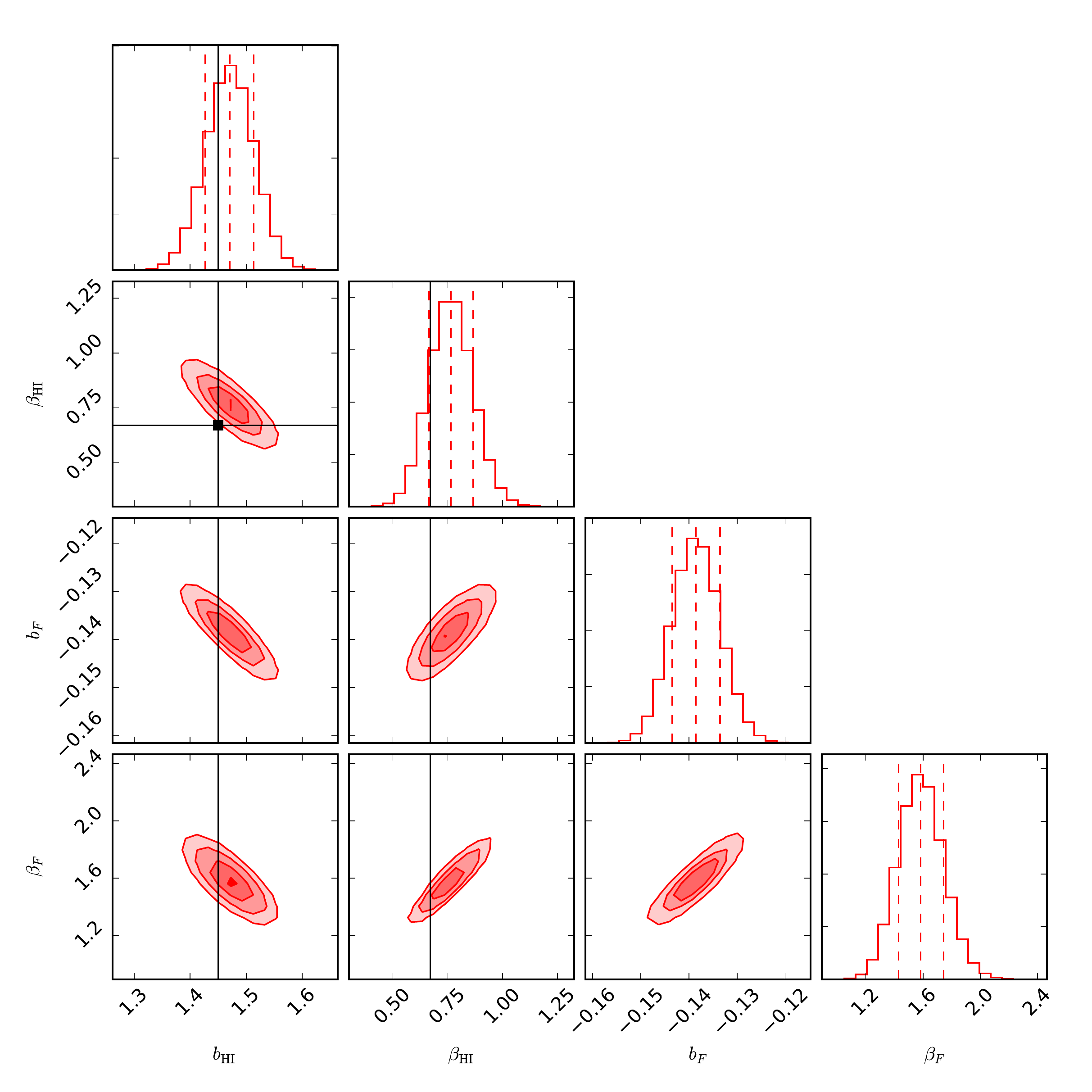}
\caption{Constraints and degeneracies on the bias parameters of the 21cm and Ly$\alpha$ forest obtained by performing a joint fit to the auto- and cross-power spectra of the two fields. The blue lines indicate the expected value for the 21cm, that we know by construction.}
\label{fig:corner}
\end{center}
\end{figure}

{\bf Method 2: cross $P(k)$ multipoles.} In this second method we exploit also the information contained in the cross-correlation power spectrum $P_{{\rm 21cm-Ly}\alpha}(\vec{k})$. Following Eqs. \ref{eq:legendre} and \ref{eq:cross_lin}, we can write the cross-power spectrum multipoles as:
\bear
\label{eq:multicross}
P_0(k)&=&A \Omega_{\rm HI} b_{\rm HI} b_F \left( 1+\f{1}{3}(\beta_F + \beta_{\rm HI} )+\f{1}{5}\beta_F \beta_{\rm HI}  \right) P_{\rm m}(k) ,\\
P_2(k) &=&A \Omega_{\rm HI} b_{\rm HI} b_F \left( \f{2}{3}(\beta_F + \beta_{\rm HI} )+\f{4}{7}\beta_F \beta_{\rm HI}  \right) P_{\rm m}(k)~.
\label{eq:multicross2}
\ear
where the normalisation factor $A$ is such that $\overline{\delta T}_b=A \Omega_{\rm HI}$ from Eq. \ref{eq:deltaTb} and the total amount of HI is set to $\Omega_{\rm HI}=10^{-3}$ in our modelling.
We fit simultaneously the monopoles and quadrupoles of the auto- and cross-power spectra, again employing the \emph{emcee} package \cite{emcee} to perform Monte Carlo Markov Chain (MCMC) on the four free parameters ($b_F$, $\beta_F$, $b_{\rm HI}$, $\beta_{\rm HI}$) employing only power spectra measurements for $k<0.2~h{\rm Mpc}^{-1}$. As in the case of the auto-power spectra, we estimate the errors on the multipoles of the cross-power spectra assuming the modes follow a Gaussian distribution and accounting for the correlation between monopoles and quadrupoles among the different auto- and cross-power spectra (see appendix \ref{app_errors}). The results are shown in the bottom row of table \ref{tab:fit} and the degeneracies among parameters are displayed in Fig. \ref{fig:corner}.

\begin{table}
\begin{center}
\begin{tabular}{|l| c|c|c|c|c|}
\hline
                  Method              & $b_F$         & $\beta_F$  & $b_{\rm HI}$ & $\beta_{\rm HI}$ & $\chi^2/{\rm dof}$  \\ \hline \hline
     auto-power spectra                & $- 0.144^{+0.007}_{-0.007}$ & $1.480^{+0.21}_{-0.20}$  & $1.520^{+0.058}_{-0.060}$ & $0.720^{+0.14}_{-0.13}$ & 13.6/12 \\ \hline
     + cross-power spectra           & $-0.139^{+0.005}_{-0.005}$  & $1.579^{+0.16}_{-0.15}$  & $1.472^{+0.043}_{-0.044}$ & $0.761^{+0.10}_{-0.10}$ & 30.2/20 \\ \hline
    \end{tabular}
    \caption{Value of the bias and $\beta$ parameters derived by carrying out fit to the results of the simulations using the auto-power spectrum multipoles alone (upper row) and making a joint fit to all auto- and cross-power spectra of the 2 fields (bottom row).}
    \label{tab:fit}
 \end{center}
\end{table}

\begin{figure}
\begin{center}
\includegraphics[width=16cm]{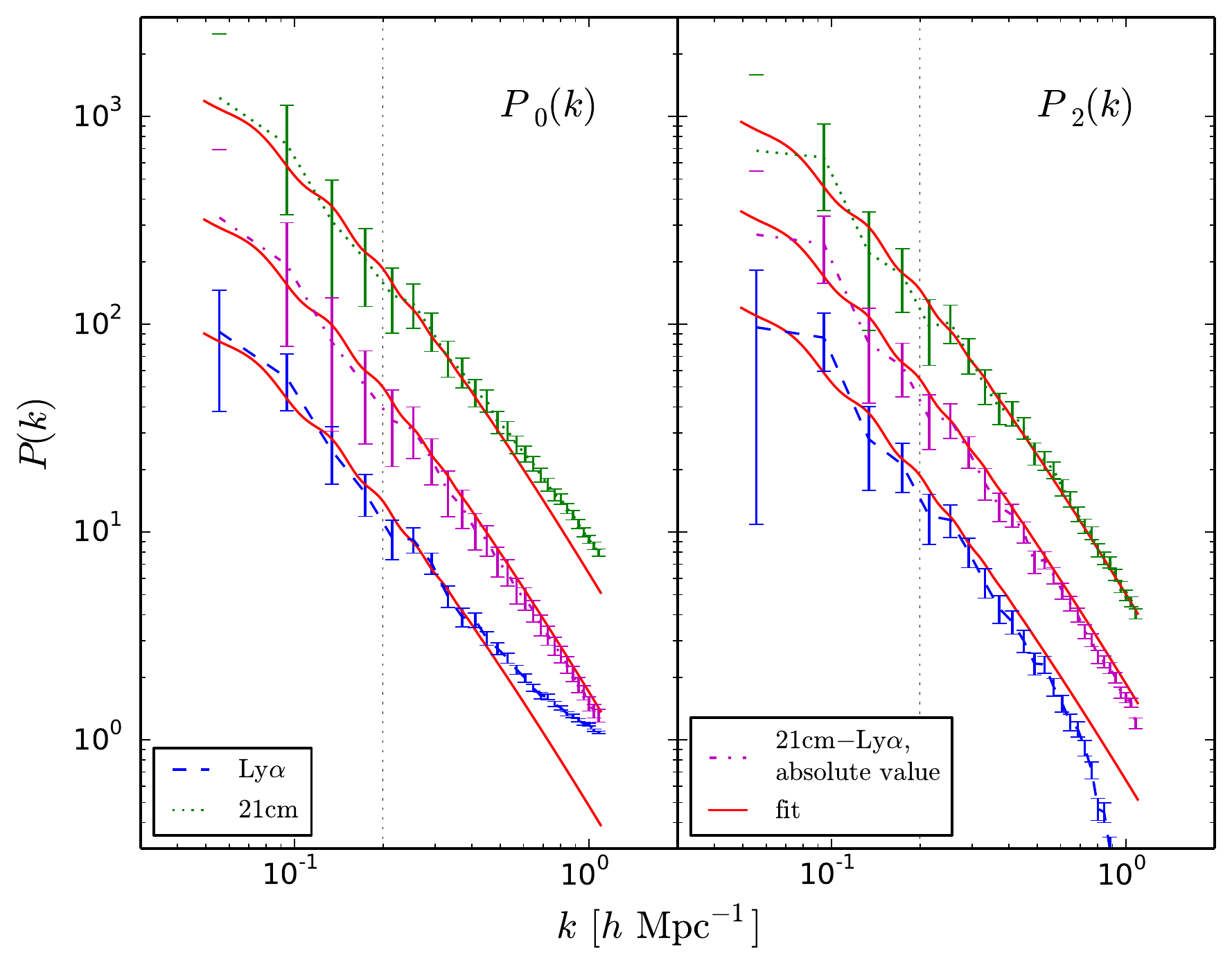}
\caption{The red solid lines are the result of fitting the $P(k)$ monopoles (left panel) and the quadrupoles (right), the green dotted lines refer to the 21cm, the blue dashed to Ly$\alpha$ forest flux and the magenta dashed-dotted to their cross-correlation in absolute value, following the analysis described in section \ref{sub:fit}. The black dotted vertical line marks $k=0.2 h \,{\rm  Mpc^{-1}}$, the mode up to which we perform the analysis.}
\label{fig:fits}
\end{center}
\end{figure}

The best-fit values for the monopoles and quadrupoles of the auto- and cross-power spectra from the joint fitting are shown in Fig. \ref{fig:fits}. For this method we obtain a normalized $\chi^2$ of 30.2/20, showing that we may be underestimating some errors. We obtain values of the parameters equal to $b_{\rm HI}=1.472^{+0.043}_{-0.044}$, $\beta_{\rm HI}=0.762^{+0.10}_{-0.10}$, $b_F=-0.139^{+0.005}_{-0.005}$, $\beta_F=1.580^{+0.16}_{-0.15}$. The derived values for the HI are in perfect agreement with the input ones, and the combination $b_F (1+\beta_F)$ also reproduces the observational constraints \cite{Blomqvist2015}. We also notice that the values derived with this method are in agreement with the ones obtained by fitting only the auto-power spectra.

\begin{figure}
\begin{center}
\includegraphics[width=12cm]{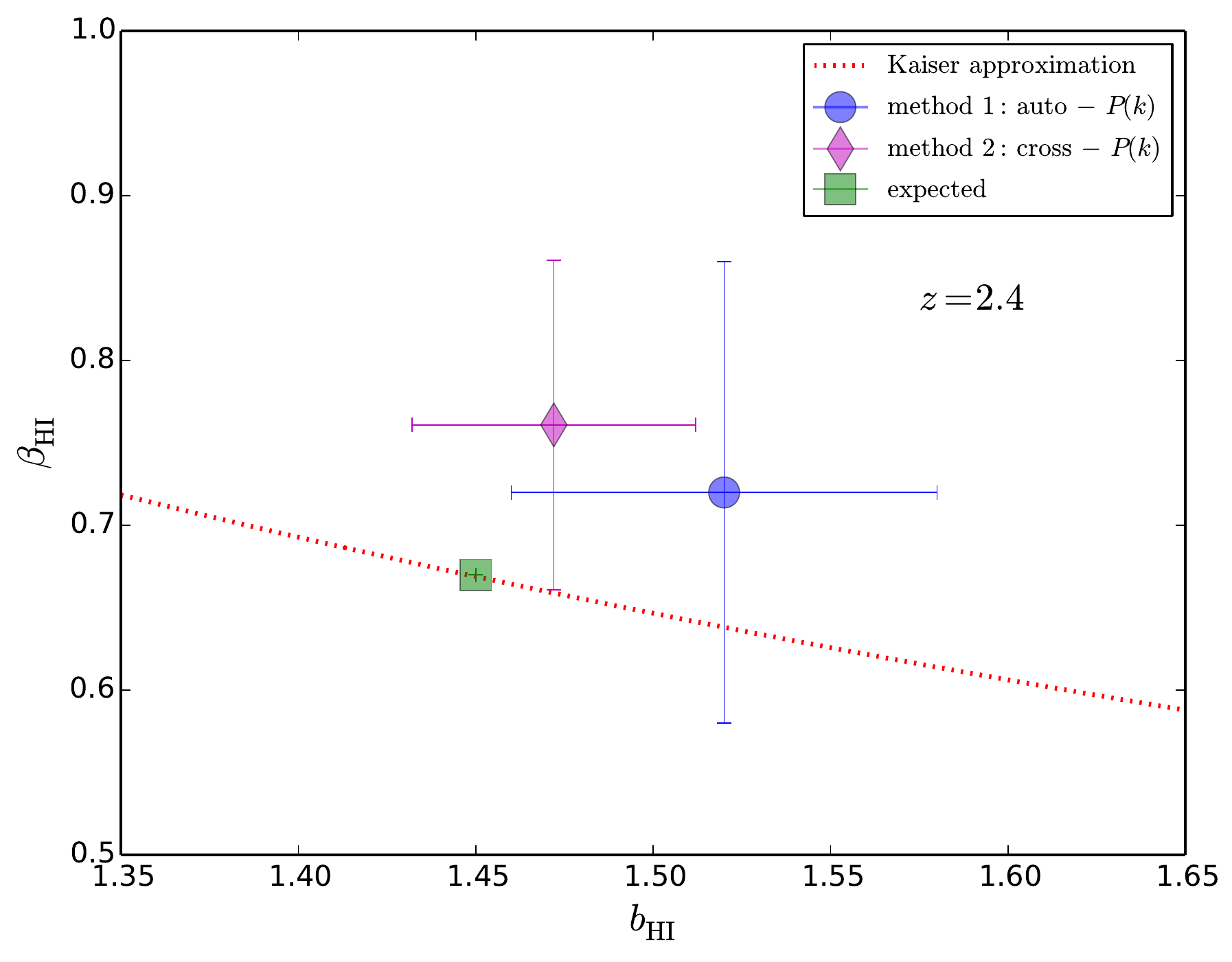}
\caption{Values of the bias parameters $b_{\rm HI}$ and $\beta_{\rm HI}$ derived with the two procedures described in section \ref{sub:fit} and summarised in table \ref{tab:fit}. The green square indicates the expected values, that we know by construction. The red dotted line is the Kaiser approximation: $\beta_{\rm HI} \times b_{\rm HI} = f (\Omega_{\mathrm m}) \simeq \Omega_{\mathrm{m}}^{0.55} (z)$.}
\label{fig:HIparam}
\end{center}
\end{figure}

We show in Fig. \ref{fig:HIparam} a more detailed comparison between the derived values of the 21cm from the two methods and the input ones. The dotted line in that plot represents the function $\beta_{\rm HI}=(f/b_{\rm HI})$, where $f(z)\simeq[\Omega_{\mathrm m}(z)]^{0.545}$ from linear theory; for the cosmological set-up employed in this simulation at redshift $z=2.4$, $f = 0.97$. There is agreement among the two $(b_{\rm HI}, \beta_{\rm HI})$ points, and both are compatible with the Kaiser approximation (dotted line) within $1\sigma$.

It is important to notice that by adding the information embedded into the cross-power spectra, together with the assumed perfectly correlation on linear scales, the errors on the 21cm and Ly$\alpha$ forest parameters decrease by $\sim30\%$. Therefore cross-power spectra have to be seen as a powerful way to look for systematics in the 21cm field but also as a way to add extra information that can shrink the error on the model parameters.

We also notice that although we are only using power spectra measurements with $k<0.2~h{\rm Mpc}^{-1}$ to fit the results of the simulations, our linear model for the cross-power spectrum, used with the best-fit value parameters from the fit, is capable of reproducing the amplitude and shape of the cross-power spectrum multipoles to a remarkable accuracy down to the smallest scales we probe: see Fig. \ref{fig:fits}. This could be just a coincidence, it may arises because we are not properly modeling the 1-halo term of the HI distribution or may be something more deep. Investigating this issue is however beyond the scope of this paper.

We want to stress again that here our task was to retrieve the input $b_{\rm HI}$ value through our simulated data as a consistency check and to show the error improvement by using the information coming from the cross-power spectrum. We do not attempt to constraint the absolute $b_{\rm HI}$ since it is tuned in our model by construction.

One possible systematic in our analysis is the impact of the HI density profile we have used to compute 21cm power spectra. In the appendix \ref{appHI} we show that on the scales where we carry out the fit, the 1-halo term does not have a noticeable effect. We also notice that using a model with a different value of $\Omega_{\rm HI}$ will only shift the amplitude of the 21cm power spectrum, but not its shape, i.e. the values of $(b_{\rm HI}, \beta_{\rm HI})$ will not be affected.

Since the amplitude of the 21cm power is proportional to $\Omega_{\rm HI}b_{\rm HI}$, 21cm intensity mapping surveys are only sensitive to that product: e.g. $\Omega_{\rm HI}b_{\rm HI} = 0.62 \times 10^{-3}$ at $z \sim 0.8$ \cite{switzer2013}; for comparison, we remind the reader that in our modelling we set $\Omega_{\rm HI}=10^{-3}$ as an input parameter. We stress that the degeneracy $\Omega_{\rm HI} - b_{\rm HI}$ can be broken by adding information from other surveys, such as the HI column density distribution function from the Ly$\alpha$ forest, that is directly sensitive to $\Omega_{\rm HI}$. From the 21cm signal alone (after reionization) the $\Omega_{\rm HI}$ measurement is expected to be hard if not impossible, as result of the drastically dropping of the brightness temperature $T_b$ (the strength of the signal), making it less easy to distinguish from foregrounds.

The effect of shifting the 21cm power spectrum amplitude can be produced also by allowing the dark matter particle to have mass of the order of keV (thermal relic) instead of the {\it cold} approximation \cite{carucci2015}, and this mass scale is, for example, that of the sterile neutrino, one of today most favourable dark matter candidates \cite{sterileNeutrino}. Hence, to extract information about $\Omega_{\rm HI}$ by looking at the observed $P_{\rm 21cm}(k)$, one should also consider the nature of the dark matter particle and disentangle the two responses on the $P_{\rm 21cm}(k)$.

\subsubsection{Degeneracies between the bias $b_{\rm HI}$, $\Omega_{\rm HI}$ and the growth factor $f$: a fisher matrix analysis}
\label{sub:fisher}

Given the results shown in Sec. \ref{sub:fit}, we can infer that the Kaiser approximation from linear theory $f=\beta_{\rm HI}b_{\rm HI}$ agrees well with our 21cm-Ly$\alpha$ modelling (Fig. \ref{fig:HIparam}) over a compelling range of $k$ (Fig. \ref{fig:fits}) and thus we can rewrite the linear order 21cm power spectrum of Eq. \ref{eq:21cm_lin} replacing $\beta_{\rm HI}=f/b_{\rm HI}$; its monopole and quadrupole become:
\bear
\label{eq:multipoles_f}
P_{{\rm 21cm},0}(k)&=&A^2 \Omega_{\rm HI}^2 \left(b_{\rm HI}^2+\f{2}{3}f b_{\rm HI}+\f{1}{5}f^2  \right) P_{\rm m}(k) ,\\
P_{{\rm 21cm},2}(k) &=&A^2 \Omega_{\rm HI}^2f  \left( \f{4}{3}b_{\rm HI} +\f{4}{7}f \right) P_{\rm m}(k)  .
\label{eq:multipoles2_f}
\ear
Analogously, the cross-power spectrum multipoles of Eqs. \ref{eq:multicross}-\ref{eq:multicross2} become:
\bear
\label{eq:multicross_f}
P_0(k)&=&A \Omega_{\rm HI} b_F \left( b_{\rm HI}+\f{1}{3}(f + b_{\rm HI}\beta_{\rm F} )+\f{1}{5} f \beta_F  \right) P_{\rm m}(k) ,\\
P_2(k) &=&A \Omega_{\rm HI}  b_F \left( \f{2}{3}(\beta_F b_{\rm HI} + f )+\f{4}{7}\beta_F f  \right) P_{\rm m}(k)~.
\ear
We thus perform a new MCMC analysis as described in Sec. \ref{sub:fit}, for determining the parameters ($b_F$, $\beta_F$, $b_{\rm HI}$, $f$) employing only power spectra measurements for $k<0.2~h{\rm Mpc}^{-1}$, first fitting together the monopoles and quadrupoles of the auto-power spectra, and then adding also the cross-power spectrum. Cosmic variance errors (no system noise) are estimated again as in appendix \ref{app_errors}. The results are shown in table \ref{tab:fit_kaiser}.

\begin{table}
	\begin{center}
		\begin{tabular}{|l| c|c|c|c|c|}
			\hline
			Method              & $b_F$         & $\beta_F$  & $b_{\rm HI}$ & $f$ & $\chi^2/{\rm dof}$  \\ \hline \hline
			auto-power spectra                & $- 0.144^{+0.007}_{-0.007}$ & $1.478^{+0.21}_{-0.20}$  & $1.515^{+0.046}_{-0.047}$ & $1.1^{+0.09}_{-0.09}$ & 24.4/12 \\ \hline
			+ cross-power spectra           & $-0.141^{+0.005}_{-0.005}$  & $1.508^{+0.18}_{-0.17}$  & $1.485^{+0.043}_{-0.044}$ & $1.1^{+0.08}_{-0.08}$ & 44.3/20 \\ \hline
		\end{tabular}
		\caption{Value of the bias and $\beta$ parameters and of the cosmological growth factor $f$ derived by carrying out fit to the results of the simulations using the auto-power spectrum multipoles alone (upper row) and making a joint fit to all auto- and cross-power spectra of the 2 fields (bottom row).}
		\label{tab:fit_kaiser}
	\end{center}
\end{table}

There is agreement between the ($b_F$, $\beta_F$, $b_{\rm HI}$, $f$) values determined with the two fits and adding information coming from the cross-power spectrum again shrinks the associated errors. Both fits prefer an unreasonable growth factor value greater than unity, $f \simeq 1.1 \pm 0.1$, although being in agreement with linear theory $f = f (\Omega_{\mathrm m}) \simeq \Omega_{\mathrm{m}}^{0.55} (z) = 0.97$ within 1$\sigma$. This can be seen already in Fig. \ref{fig:HIparam} where both points were above the dotted line corresponding to the linear relation $\beta_{\rm HI} \times b_{\rm HI} = f (\Omega_{\mathrm m})$. We remind that we do not add any prior on the physical value of $f$.

To better understand the degree of degeneracy of parameters involved in the 21cm characterization ($\Omega_{\rm HI}$, $b_{\rm HI}$ and $f$), we perform a Fisher matrix analysis using monopoles and quadrupoles of the 21cm auto-power spectrum and the of cross-power spectrum with the \lya flux. This is a good exercise especially to check the effect of the uncertainty of $\Omega_{\rm HI}$, that is tuned by construction in the previous analysis and in the simulated 21cm field.

The Fisher matrix analysis quantifies the amount of information that the 21 cm power spectrum as observable carries about the three parameters $\Omega_{\rm HI}$ - $b_{\rm HI}$ - $f$. Practically, we use as prior the values found with the MCMC fit as in table \ref{tab:fit_kaiser} (i.e. $\Omega_{\rm HI} =10^{-3}$, $b_{\rm HI} = 1.5$ and $f = 1.1$), and we calculate analytically how much the 21cm power spectrum varies by varying the values of those parameters. We make use of Eqs. \ref{eq:multipoles}-\ref{eq:multipoles2}-\ref{eq:multicross}-\ref{eq:multicross2} as templates and of the expressions shown in the appendix \ref{app_errors} for building the covariance matrix, i.e. the Gaussian uncertainties linked to these parameters. A quick and clear reference for this kind of analysis is in \cite{fisher_ref}.

\begin{figure}
	
	\begin{subfigure}{0.5\textwidth}
		\includegraphics[width=7.5cm]{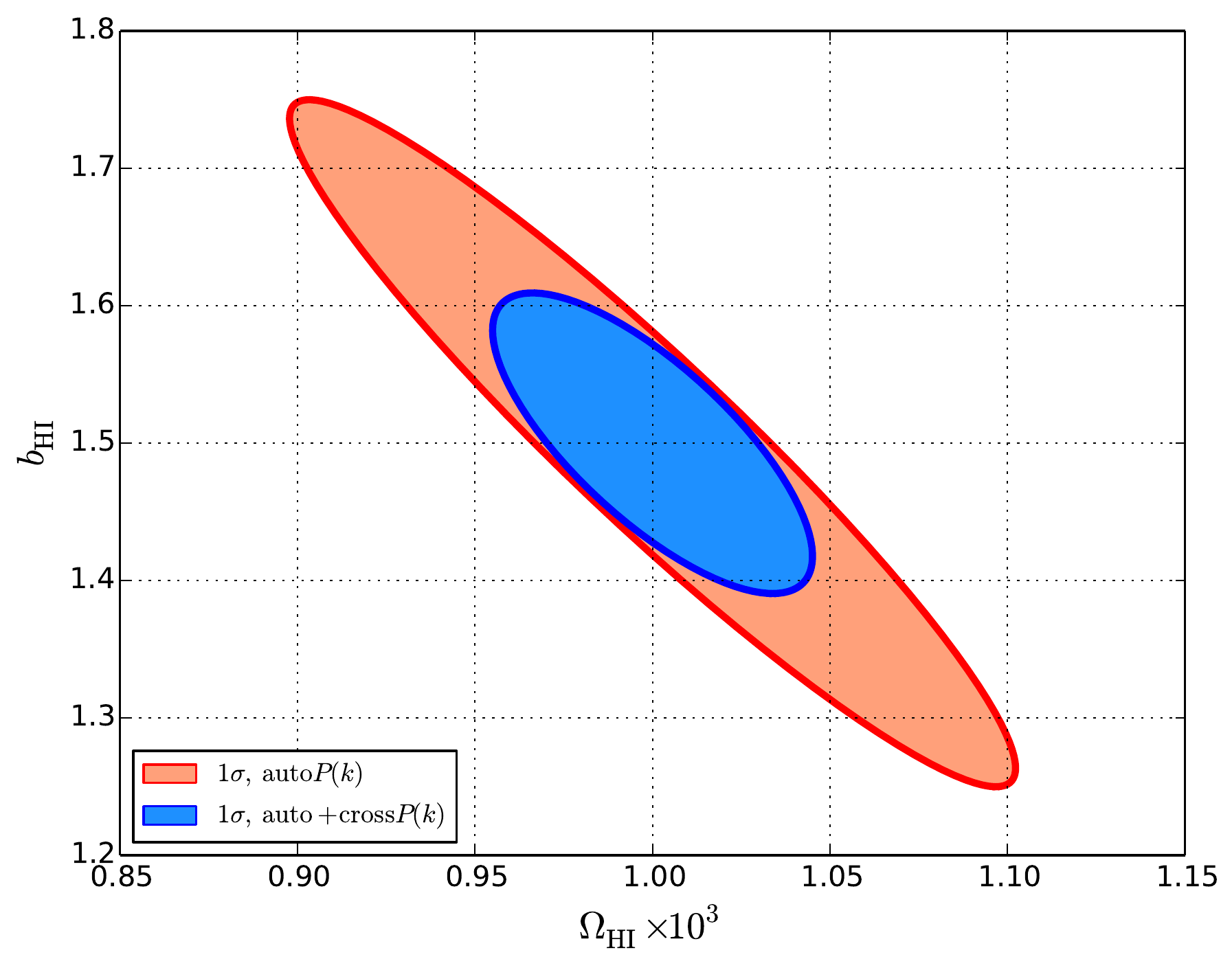} 
	\end{subfigure}
	\begin{subfigure}{0.5\textwidth}
		\includegraphics[width=7.5cm]{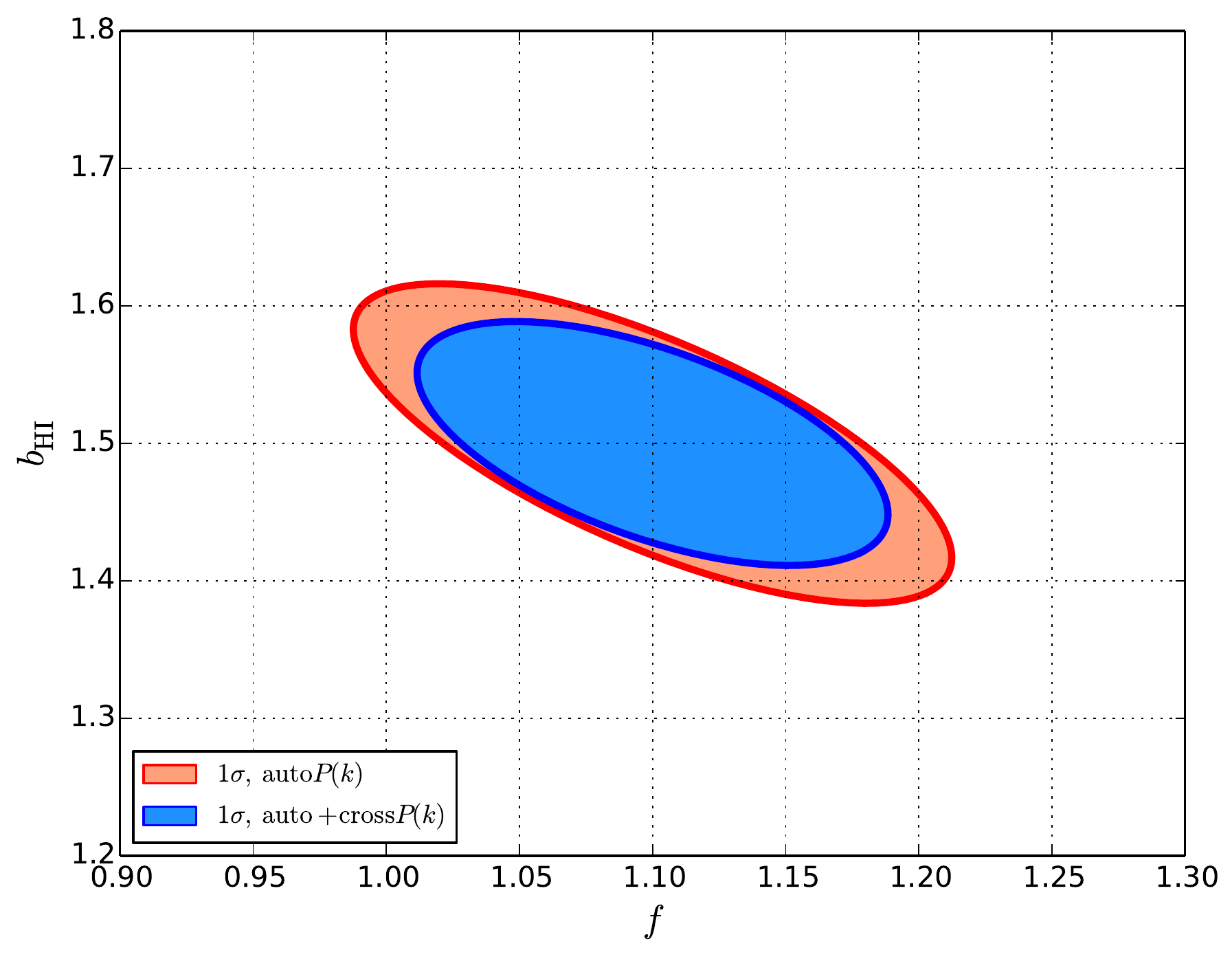} 
	\end{subfigure}

	\begin{subfigure}{0.5\textwidth}
		\includegraphics[width=7.5cm]{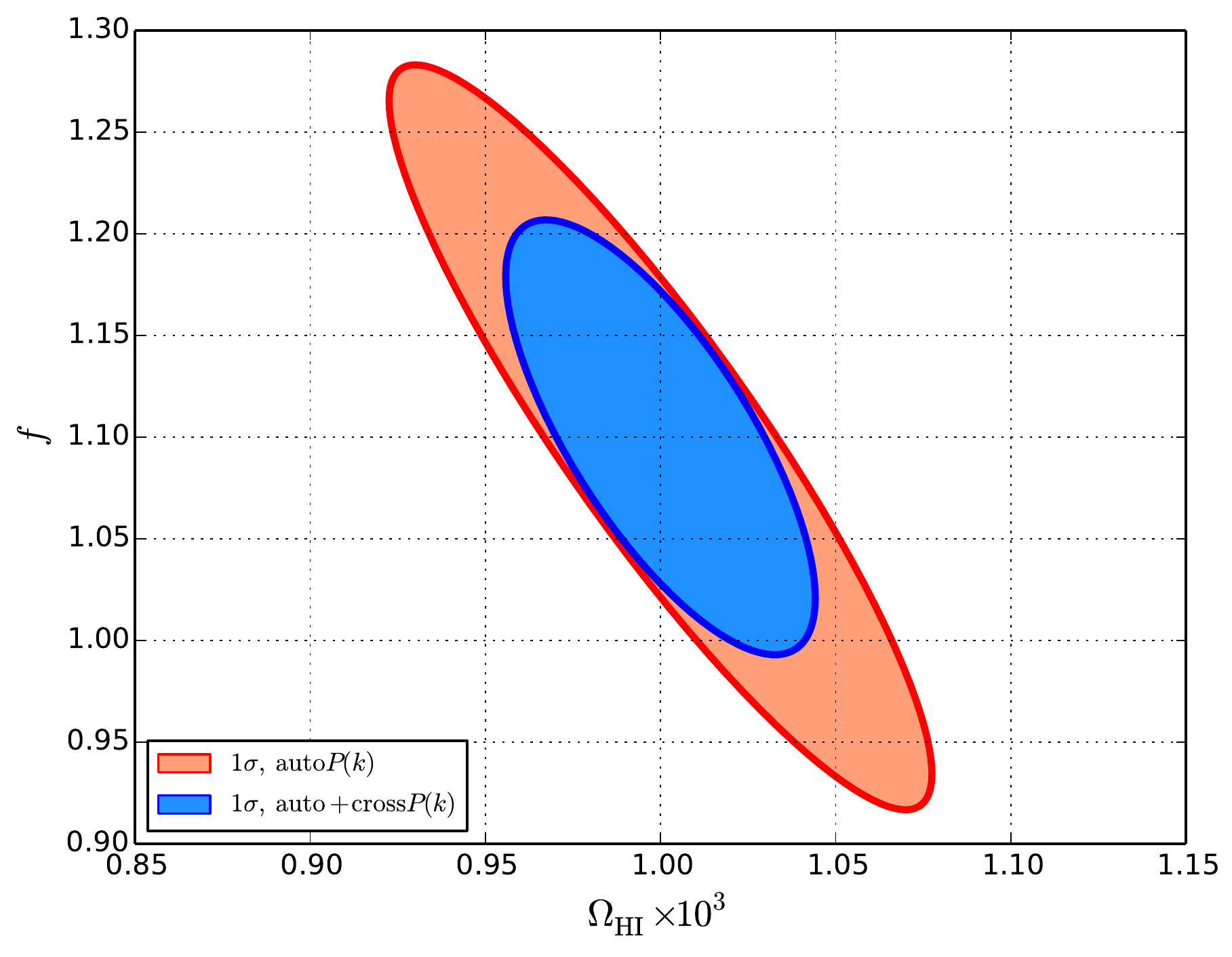} 
	\end{subfigure}
	\begin{subfigure}{0.5\textwidth}
		\includegraphics[width=7.5cm]{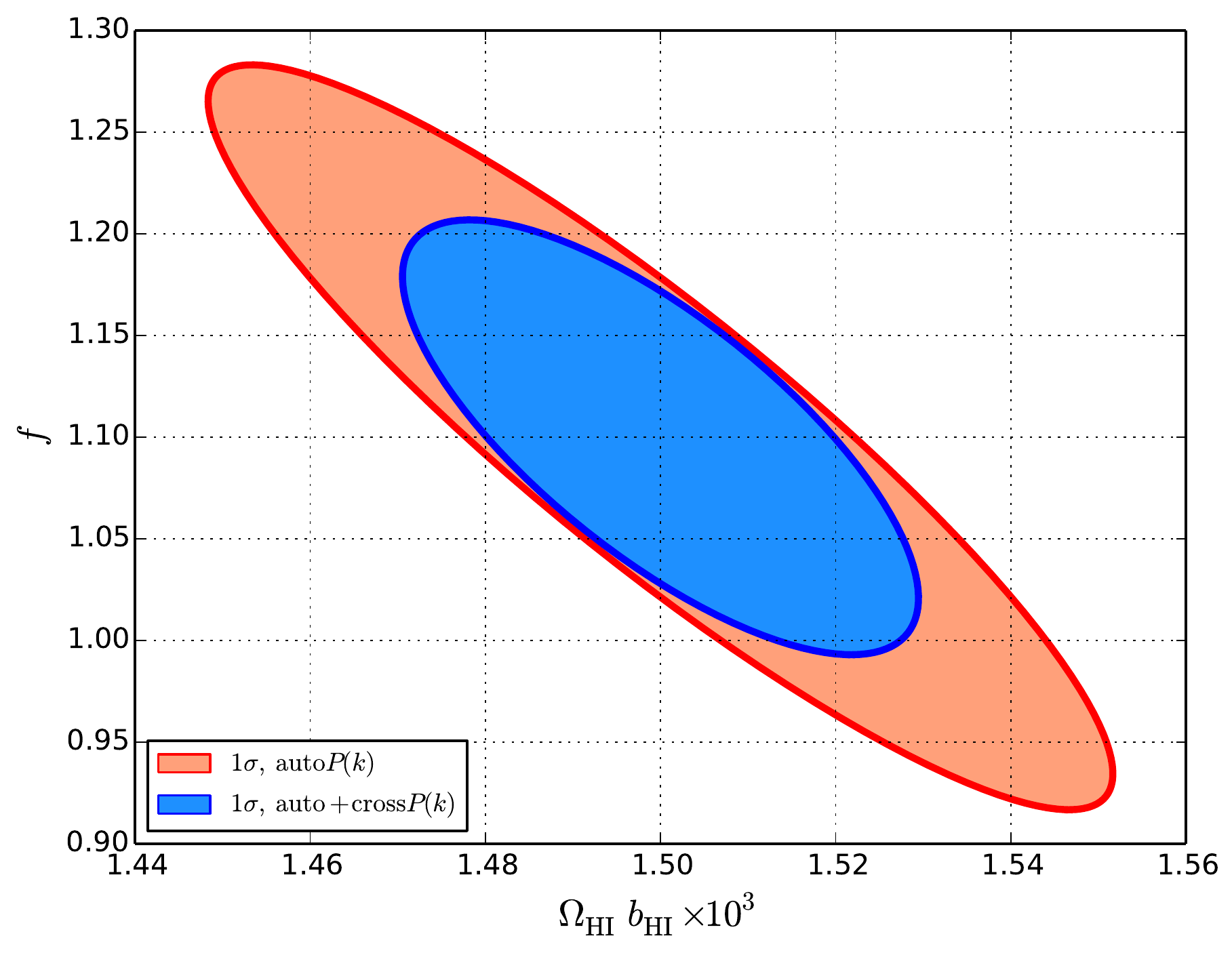} 
	\end{subfigure}
	
	\caption{1$\sigma$ contours of the values of the bias $b_{\rm HI}$, the density parameter $\Omega_{\rm HI}$ and the growth factor $f$ determined using either only the 21cm power spectrum (orange areas) or adding the cross 21cm-Ly$\alpha$ power spectrum (blue areas). The Fisher matrix analysis is performed using the theoretical templates of the power spectra multipoles and errors described in the results Sec. \ref{sec:results} and in the appendix \ref{app_errors}, using the $k<0.2 h \,{\rm  Mpc^{-1}}$ information.}
	\label{fig:fisher}
\end{figure}

In Fig. \ref{fig:fisher} we show the results. In the top left panel we show the uncertainty contours for $\Omega_{\rm HI}$ - $b_{\rm HI}$ (fixing $f$), top right the uncertainty contours for $f$ - $b_{\rm HI}$ (fixing $\Omega_{\rm HI}$), bottom left the uncertainty contours for $\Omega_{\rm HI}$ - $f$ (fixing $b_{\rm HI}$) and in last panel on bottom left the uncertainty contours for the product $\Omega_{\rm HI}$  $b_{\rm HI}$ - $f$. The orange ellipses refer to the uncertainty using the auto-power spectrum information only, the blue ellipses using information coming from both auto- and cross- power spectra: in the second case the uncertainties are always reduced, i.e. the cross-power spectrum shrinks our constraints in any case. Especially for the $\Omega_{\rm HI}$ - $b_{\rm HI}$ correlation: fixing the growth factor helps reducing the degeneracy and adding the cross- information shrinks the errors by $\sim 50\%$.

Concerning the ability of a 21cm - \lya joint analysis to constrain the cosmological growth factor $f$: we will need an independent measurement of $\Omega_{\rm HI}$ in order to reach sufficient precision. Anyway, we point out that in the redshift range probed by 21cm intensity mapping surveys and \lya flux experiments ($z\sim2-3$) we have no other precise $f$ measurement.

\section{Summary and conclusions}
\label{sec:conclusions}

The spatial distribution of matter in the Universe embeds a huge amount of information on the fraction that each component contributes to the total energy content of the Universe, on the nature of gravity and of the initial conditions of the Universe, on the geometry of the Universe and so on. Unfortunately, the spatial distribution of matter is not directly observable, but can be mapped through tracers of it such as galaxies or cosmic neutral hydrogen. 

The 21cm intensity mapping technique consists in carrying out low angular resolution radio-observations with the goal of measuring 21cm flux perturbations from large patches of the sky where the galaxies that host the HI are not resolved. It is expected that this technique will play a major role in cosmology, given the spectroscopic nature of these observations and the large volumes it can sample. Unfortunately, the cosmological signal is buried by the galactic and extragalactic foregrounds, whose amplitude is several order of magnitude larger. Therefore, the cosmological information that can be extracted from these surveys critically depends on the precision with which the foregrounds can be cleaned from the 21cm maps.

While  \textit{well behaved} foregrounds can be robustly removed, the presence of some foregrounds such as polarized synchrotron radiation may not be completely removed and can ruin the inferred shape and amplitude of the 21cm power spectrum \cite{Alonso_2015, villaescusa2015}. A way to get rid of this problem is through cross-correlations \cite{villaescusa2015, Sobacchi_2016,Wolz_2016, Lidz_2009,sarkar,Sarkar_2016}, since the foregrounds should not be correlated with the cosmological signal from the different tracers. 

In this paper we have studied the cross-correlation between 21cm maps and the Ly$\alpha$ forest at redshift $z=2.4$. We performed this study by means of state-of-the-art hydrodynamic simulations: the Sherwood suite \cite{bolton2016}. While the properties of the Ly$\alpha$ forest are obtained directly from the simulation, we use a simple deterministic model to model the abundance of HI. We compute auto- and cross-power spectra for the two different fields in redshift space. 

We find that on large, linear scales, the Ly$\alpha$ forest is completely anti-correlated ($r=-1$) with the 21cm field. This happens because regions with large HI densities exhibit large 21cm emission, but those regions are dominated by damped Ly$\alpha$ systems (DLAs), i.e. the mean transmitted flux is low. The fact that 
the Ly$\alpha$ forest traces low-density, highly ionized gas, while the 21cm signal arises from high density regions where hydrogen is self-shielded and therefore mainly neutral is the origin of this anti-correlation. 

We have computed the cross-correlation coefficient between the two fields and we find it to be -1 until $k\simeq0.2~h{\rm Mpc}^{-1}$, while on smaller scales its value increases, showing that on those scales the two fields are not perfectly anti-correlated. We thus determine that a joint fit to the different auto- and cross-power spectra using linear theory as template should be performed until $k=0.2~h{\rm Mpc}^{-1}$. This is important for fisher matrix analysis, like the one in \cite{sarkar}, used to forecast the improvement on the cosmological parameters from future surveys.

We compute errors on the monopoles of the Ly$\alpha$ forest and 21cm auto-power spectra and on the monopole of the cross-power spectrum for a cosmological volume equal to the one jointly sampled by a BOSS like survey and by SKA1-MID by means of interferometry observations. We find that the cross-power spectrum can be measured with a high S/N ratio to very small scales: S/N$>3$ for $k\in[0.06-1]~h{\rm Mpc}^{-1}$. 

We have exploited the anisotropy of the power spectra in redshift-space to determine the values of the bias and redshift-space distortion parameter, $\beta$, for both fields by computing the monopole and quadrupole for each auto- and cross-power spectrum. We have fit simultaneously the monopole and quadrupole of each field using the linear theory prediction up to the above $k_{\rm max}=0.2~h{\rm Mpc}^{-1}$ value. The errors on the multipoles are computed assuming the fields are Gaussian and the covariance is built taken into account the correlation between the monopole and quadrupole. We find that we are able to retrieve the input value for the HI distribution of $b_{\rm HI}$ together with $\beta_{\rm HI}=f/b_{\rm HI}$. In terms of the Ly$\alpha$ forest we obtain values of $\beta_{\rm F}$ and $b_{\rm F}$ that reproduce the observational constraints: $\beta_{\rm F}=1.39 \pm 0.1$ and $b_{\rm F} (1+\beta_{\rm F})=-0.374 \pm 0.007$ \cite{Blomqvist2015}. 

We also perform a joint fitting of the multipoles and quadrupoles of the Ly$\alpha$ and 21cm auto-power spectra and their cross-power spectrum. In this case the values we obtain for the bias parameters are in agreement with those derived from the auto-power spectra, but their errors are smaller by $\sim30\%$. 

We find that the multipoles of the 21cm-Ly$\alpha$ cross-power spectrum can be surprisingly well reproduced by linear theory down to the smallest scales we probe in our analysis: $k\simeq1~h{\rm Mpc}^{-1}$. It is beyond the scope of this paper to investigate whether this is just a coincidence or the result of some non-trivial physical reason.

Finally, we performed a Fisher matrix analysis for inspecting the degeneracy of the parameters characterizing the 21cm power spectrum, again showing how adding the information from the cross-correlation helps reducing the uncertainties.

In summary: while residual foreground contamination can bias the shape and amplitude of the 21cm auto-power spectrum, and therefore the estimated value of the cosmological and astrophysical parameters, the cross-power spectrum will be less affected by this problem and therefore represents a powerful method to verify the presence of systematics in the 21cm maps. If the 21cm maps are free of large systematics, data from auto- and cross-power spectra can be combined to break degeneracies and to tighten the value of the parameters. Determining the scale where linear theory holds is important since using an incorrect theoretical template will bias the inferred values of the model parameters.

\section*{Acknowledgements}
We thank Jingjing Shi, Giulio Fabbian, Andrej Obuljen, Justin Alsing, Stephen Feeney and Sigurd Naess for useful discussions. The hydrodynamical simulations used in this work were performed with supercomputer time awarded by the Partnership for Advanced Computing in Europe (PRACE) 8th Call. We acknowledge PRACE for awarding us access to the Curie supercomputer, based in France at the Tr\`es Grand Centre de Calcul (TGCC). This work also made use of the DiRAC High Performance Computing System (HPCS) and the COSMOS shared memory service at the University of Cambridge. These are operated on behalf of the STFC DiRAC HPC facility. This equipment is funded by BIS National E-infrastructure capital grant ST/J005673/1 and STFC grants ST/H008586/1, ST/K00333X/1.

\appendix
\section{21cm power spectrum: dependence on $\rho_{\rm HI}(r)$}
\label{appHI}

Here, we investigate the dependence of the HI and 21cm power spectra on the model used for the HI density profile within halos: $\rho_{\rm HI}(r|M,z)$. We study this using an hydrodynamic simulation that we coin 60-512. As the simulations used in this work, the 60-512 has been run using the TreePM+SPH code Gadget-3, with cosmological parameters in agreement with recent Planck data \citep{planck15}: ($\Omega_{\rm m}$,  $\Omega_{\Lambda}$,  $\Omega_{\rm b}$,  $h$, $n_s$,  $\sigma_8$) have the following values: (0.3175, 0.6825, 0.049, 0.671, 0.9624, 0.834).

\begin{figure}
\begin{center}
\includegraphics[width=12cm]{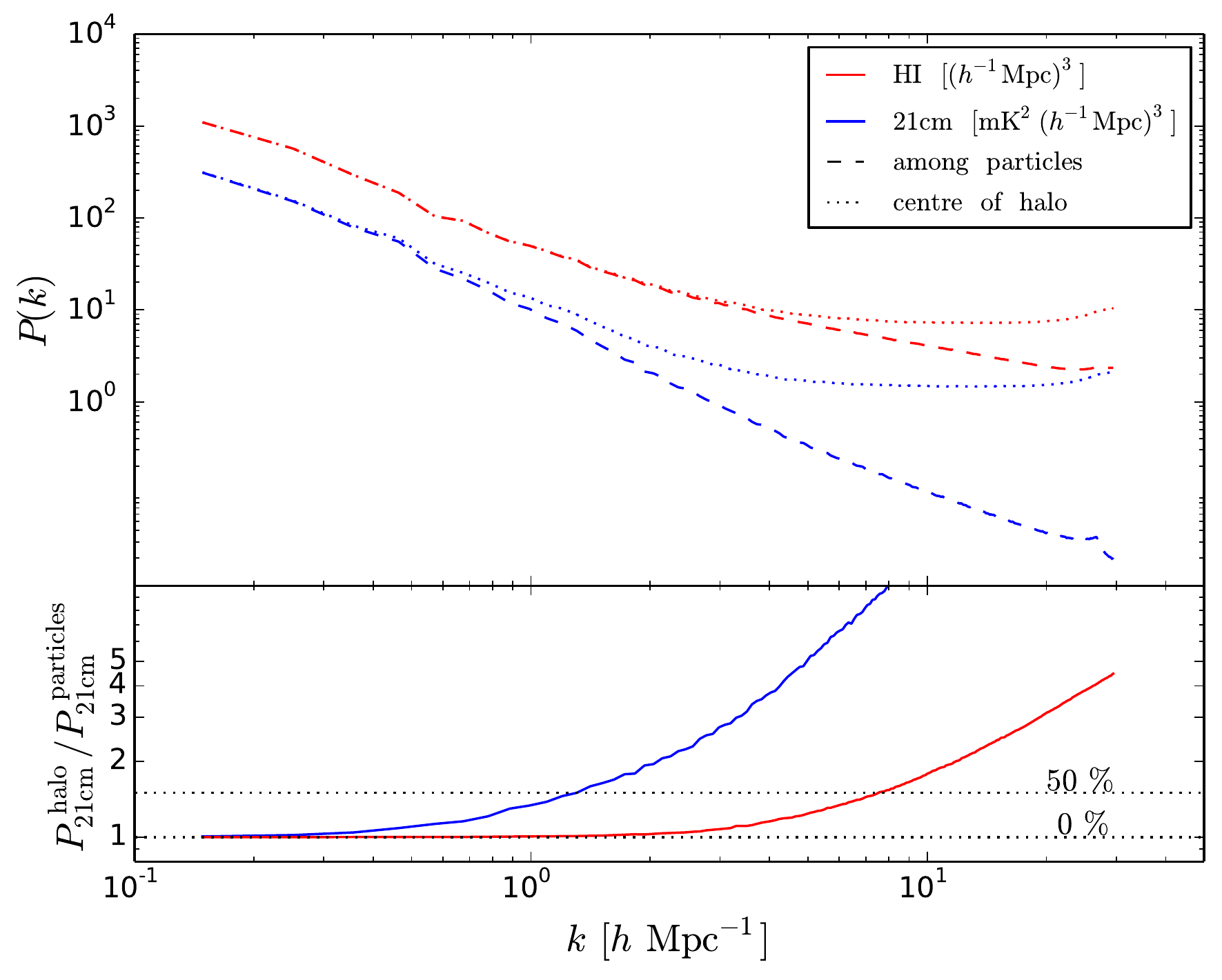}
\caption{{\it Upper panel:} The HI power spectrum in real space $P_{\rm HI}(k)$ (in red) and the 21cm $P_{\rm 21cm}(k)$ (in blue), using two ways for distributing the HI: at the centre of each halo (dotted lines) or spread among all gas particles belonging to the halo (dashed lines). {\it Lower panel:} Ratio between the two $P(k)$ computed with the HI at the centre of halos over the one with HI among all halo particles. The dotted horizontal lines mark when a $50 \%$ difference among the two $P(k)$'s is reached and when there is no difference among the two. For a description of the two ways of implementing the halo-based model see subsection \ref{sub:bagla}. All power spectra here refer to the 60-512 simulation.}
\label{fig:bagla}
\end{center}
\end{figure}

The 60-512 simulation follows the evolution of $512^3$ cold dark matter particles and $512^3$ baryon particles within a periodic box of linear comoving size of 60 $h^{-1}$ Mpc from $z=99$ down to $z=3$. Star formation is modeled using the effective multi-phase model of Springel \& Hernquist \cite{Springel-Hernquist_2003}. The code also simulates radiative cooling by hydrogen and helium and heating by an uniform Ultraviolet (UV) background. 60-512 has mass resolutions of $m_{\rm CDM} = 1.2\times10^8 \,h^{-1}{\rm M}_{\odot}$ and $m_{\rm baryon} = 2.2\times10^7\, h^{-1}{\rm M}_{\odot}$ and softening length with value $l_{\rm soft} =2.9\,h^{-1} $kpc.

We have performed our tests using this simulation and not 80-2048 or 160-1048 because these are affected by an unrealistic star formation rate that make the gas reservoir of those simulations unreliable. In our analysis we have modeled the HI density profile as:
\be
\rho_{\rm HI}(r|M,z)=M_{\rm HI}(M,z)\delta(\vec{r})
\label{eq:rho_HI}
\ee
where $\delta(\vec{x})$ is the Dirac delta. Thus, we are collapsing the HI density distribution into a single point located in the halo center. This is of course a very crude approximation, but given the fact that in our simulations the spatial distribution of gas is unreliable this is the most simple choice that does not involve ad-hoc assumptions. In order to check the dependence of our results on $\rho_{\rm HI}(r|M,z)$ we have also distributed the HI within halos evenly among all gas particles belonging to it, i.e. we have used $\rho_{\rm HI}(r|M,z)=\rho_{\rm g}(r|M,z)$, where $\rho_{\rm g}(r|M,z)$ represents the density profile of gas within a dark matter halo of mass $M$ at redshift $z$.

In Fig. \ref{fig:bagla} we show the HI and 21cm power spectra that we obtain using the two different HI density profiles. As expected, the HI density profile only affects the 1-halo term and therefore differences among different models for the $\rho_{\rm HI}(r|M,z)$ only show up on relatively small scales. On the other hand, dispersion velocities within halos propagate into large scales through redshift-space distortions; we find that the 21cm power spectra start deviating from each other on scales $k\lesssim0.03~h{\rm Mpc}^{-1}$, and differ by a $\sim50\%$ at $k\sim1.5~h{\rm Mpc}^{-1}$. These differences are thus expected since with our fiducial collapsed HI density profile we are not modeling the finger-of-God because no halo substructure is used to place the neutral hydrogen.

In this paper we are interested on the amplitude and shape of the auto- and cross-power spectra on large-scales, where the different models for the HI density profile produce almost identical results. Thus, we conclude that our findings are robust against the simplified model we use to distribute HI within halos.

\section{On some standard practices in generating mock Ly$\alpha$ forest spectra}
\label{app}

In this appendix we discuss some of the problems we have faced when computing the 3D power spectrum of the Ly$\alpha$ forest. As in appendix \ref{appHI}, here we also make use of the 60-512 simulation. We extract mock Ly$\alpha$ absorption spectra skewers from it as described in section \ref{sec:Lya}, i.e. in the same way as the spectra analysed in the whole paper. All absorption spectra used in this appendix contain 256 pixels each and are taken along the $x$ direction of the 60-512 simulation box.

\subsection{Normalising the spectra with $\tau_{\rm eff}$}
\label{sec:tau}

The amplitude of the 3D Ly$\alpha$ power spectrum, $P_{{\rm Ly}\alpha}(k)$, depends on the actual observed cosmological mean transmitted flux $\langle F \rangle_{\rm obs}$, measured in e.g. \cite{kim2007,faucher2008,becker2013}. When the absorption spectra are recovered artificially by piercing the simulation box with skewers, the mean value $\langle F \rangle_{\rm cat}$ of the catalogue of spectra can vary quite a lot due primarily to the number of skewers that are drawn, as already seen earlier in Fig. \ref{fig:3dlyaPk}, i.e. the more we sample the box, the more flux we get and $\langle F \rangle_{\rm cat}$ increases. Of course in a survey we have a finite amount of l.o.s. ($nlos$). The question is: how should the mean flux be computed to get sensible results for the 3D Ly$\alpha$ power spectrum?

A possibility is to take a real survey l.o.s. density, for example we know that BOSS roughly detects 15 quasars' spectra per deg$^{2}$  \cite{slosar2011}, but this would make a very low number of skewers in a typical hydrodynamical simulation box and let arise other computational problems (see next subsection \ref{sec:grid}).

It is also possible to take the $nlos$ necessary to have $\langle F \rangle_{\rm cat} = \langle F \rangle_{\rm obs}$, i.e. to match the simulation flux mean with the measured one, so it would be naively expected that the simulation $P_{{\rm Ly}\alpha}(k)$ converges with the measured one. But this is not completely true as we now discuss.

\begin{figure}
\centering
\includegraphics[width=12cm]{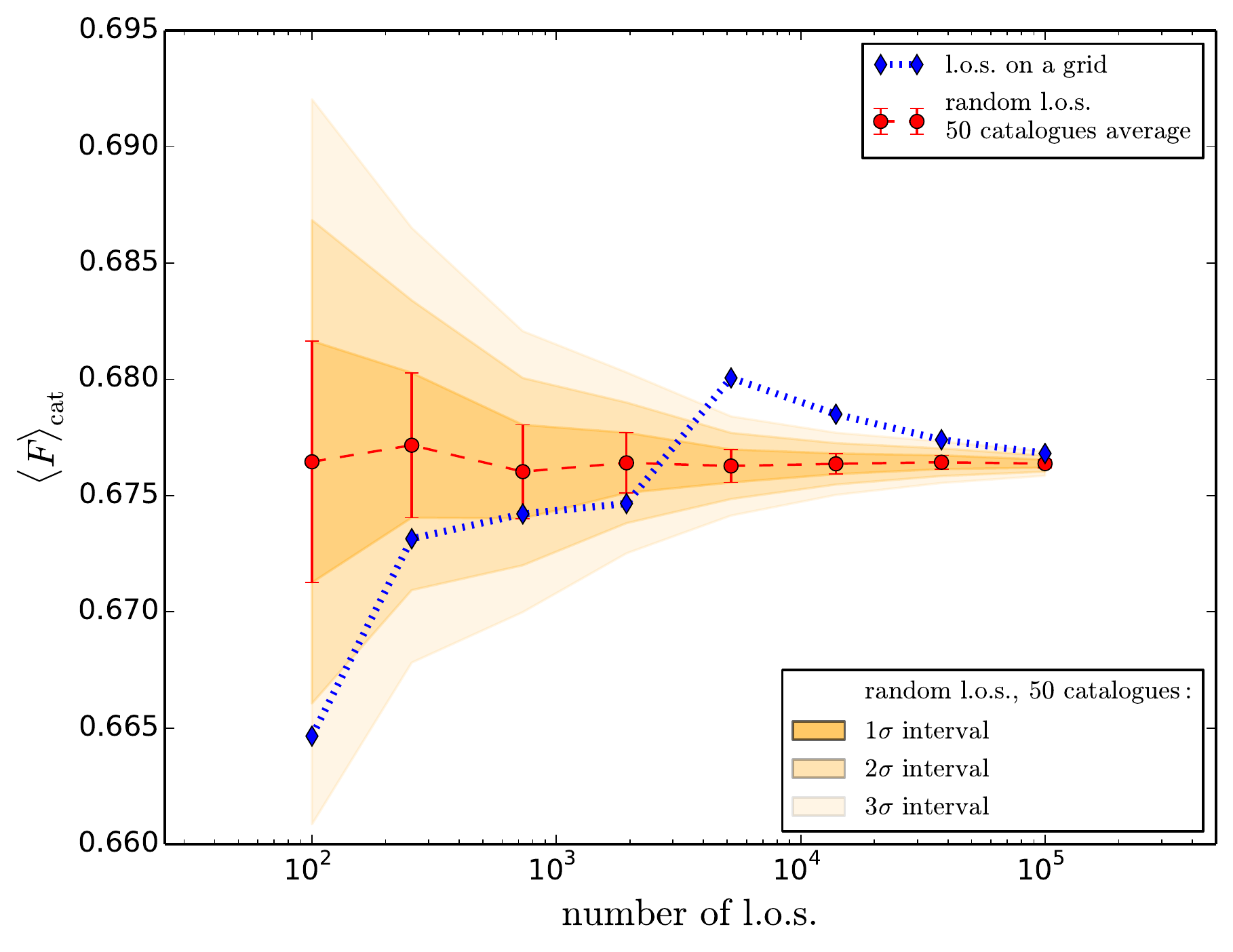}
\caption{Mean flux $\langle F \rangle$ of the Ly$\alpha$ absorption spectra catalogues, plotted versus the number of lines of sight (l.o.s.) in each catalogue. 8 catalogues have the l.o.s. placed on a regular grid (blue diamonds). Each red point represents the average of the mean flux from 50 independent catalogues with same number of random l.o.s.. The error bars display the $1\sigma$ variation from the catalogues, while the shaded orange contours show the $1\sigma$, $2\sigma$ and $3\sigma$ deviations.  All absorption spectra are made of 256 pixels skewers along the $x$ direction from  the 60-512 simulation.}
\label{fig:Fmean}
\end{figure}

\begin{figure}
\centering
\includegraphics[width=12cm]{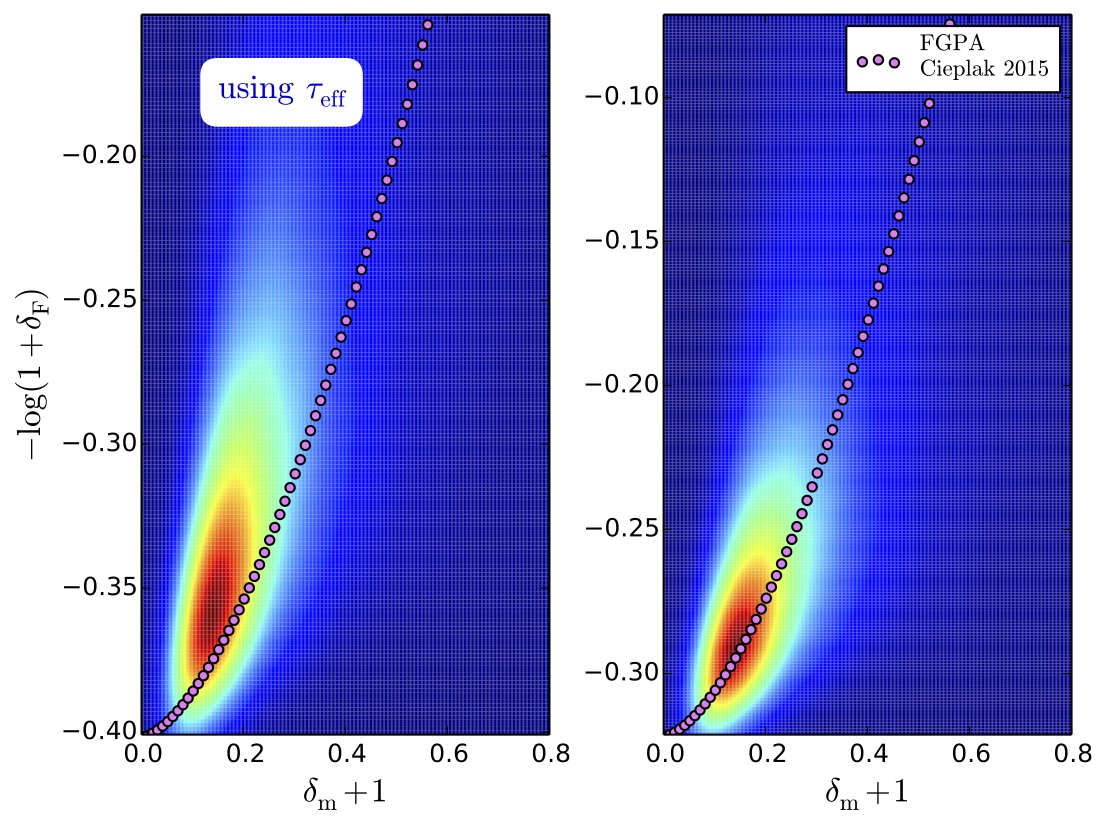}
\caption{Density of points (red color for high density values, blue for small) for the scatter of $(\delta_{\rm m} + 1)$ - $(-\log(1+\delta_{\rm F}))$ calculated for the 60-512 simulation pierced by 100 random lines of sight with 256 pixels each, as described in the subsection \ref{sec:tau}. The magenta dotted line is the fluctuating Gunn-Peterson approximation (FGPA) relation of equation \ref{eq:FGPA} with the parameters in \cite{cieplak2016}. {\it Left panel:} The spectra have been normalised by using an effective optical depth $\tau_{\rm eff}$. {\it Right panel:} The spectra have not been normalised. By eye it is clear that the right panel (unnormalised spectra) has a much better agreement with FGPA.}
\label{fig:FGPAscatterplot}
\end{figure}

In Fig. \ref{fig:Fmean} we show $\langle F \rangle_{\rm cat}$ coming from catalogues with the following $nlos$: $10^2$, $16^2$, $27^2$, $44^2$, $72^2$, $118^2$, $194^2$ and $316^2$. The skewers have been placed either on a $\sqrt{nlos}\times\sqrt{nlos}$ regular grid (blue diamonds, each point corresponds to one catalogue) or randomly (red dot, each point correspond to the average among 50 different catalogues). In case of random l.o.s., we also plot the $1\sigma$ error bar on the average $\langle F \rangle_{\rm cat}$ and show with shaded area the $2\sigma$ and $3\sigma$ intervals too. The scatter is big for the catalogues with smaller $nlos$, nevertheless their average $\langle F \rangle_{\rm cat}$ agrees with the what we obtain with a huge amount of l.o.s. We thus conclude that the catalogue mean flux is not very sensitive to the value of the surface density with which the Ly$\alpha$ field is sampled through skewers, i.e. the quantity $\langle F \rangle_{\rm cat}$ is a property of the specific realization itself, marginally dependent on the choice for $nlos$.

Throughout the literature (e.g. \cite{theuns1998,white2010,becker2013,lukic2015,bolton2016}) the general practice consisted in artificially change the amplitude of the UV background strength to obtain the desired mean transmitted flux:
\be
\langle e^{-B\tau_i} \rangle = \langle e^{-\tau_{\rm eff}} \rangle = \langle F \rangle_{\rm obs}\,,
\ee
where $\tau_i$ is the optical depth of a single pixel, $B$ is a scalar representing the variation in the amplitude of the UV background and $\tau_{\rm eff}$ is called effective optical depth (e.g. \cite{becker2013}).

By construction, employing $\tau_{\rm eff}$ makes $\langle F \rangle_{\rm cat} = \langle F \rangle_{\rm obs}$, but the shape and amplitude of the Ly$\alpha$ power spectrum is affected by this change, particularly on small scales and not only: by linearly shifting the pixel optical depth, we are non-linearly changing its flux.

To clarify this point, we make use of the fluctuating Gunn-Peterson approximation (FGPA, \cite{croft1998}) that relates the matter and transmitted flux fields through:
\be
F = e^{-A(1+\delta_{\rm m})^{\alpha}},
\label{eq:FGPA}
\ee
where $\delta_{\rm m}$ is the matter density contrast and for the other parameters we adopt the values in \cite{cieplak2016}: $A=0.3((1+z)/(1+2.4))^{4.5}$ and $\alpha=1.6$ . The FGPA, although neglecting some small scales physics, has been first derived analytically and has had great success in explaining statistical observed properties of the Ly$\alpha$ forest. We check whether FGPA holds for our mock spectra. We generate two catalogues composed by the same 100 random l.o.s. (i.e. BOSS resolution in the $60^2\,(h^{-1} {\rm Mpc})^2$ simulation box) with 256 pixels each, but one of them is normalised to $\langle F \rangle_{\rm obs}$ using $\tau_{\rm eff}$. We interpolate the flux $F$ pixels and matter particles back onto a $256^3$ grid to obtained two 3D density maps. In Fig. \ref{fig:FGPAscatterplot} in color code we display the density of points in the Ly$\alpha$ flux $F$ $-$ matter scatter plot and we plot with magenta dots the FGPA relation. The left panel refers to the normalised catalogue, the right panel to the unnormalised. We find that the flux renormalization shifts the whole flux field (see the different $y$ scale on the two panels) and modifies the FGPA relation.

To have convergence between a mock Ly$\alpha$ catalogue $P_{{\rm Ly}\alpha}(k)$ and observations or to compare different simulations Ly$\alpha$ power spectra, instead of employing the effective $\tau_{\rm eff}$, a wiser choice would be to shift whole spectra a posteriori, rather than degrade the small scale information of the flux.

None of the spectra catalogues of this work has been normalized using $\tau_{\rm eff}$, at the expenses of having the tiny (and understood) large scale discrepancies in $P_{{\rm Ly}\alpha}(k)$ (see Fig. \ref{fig:3dlyaPk}); still, the statistical properties of the Ly$\alpha$ forest flux have been remarkably recovered (see bias and redshift space distortion parameter determination in section \ref{sub:fit}).

\subsection{Placing lines of sight on a regular grid}
\label{sec:grid}

\begin{figure}
\centering
\includegraphics[width=12cm]{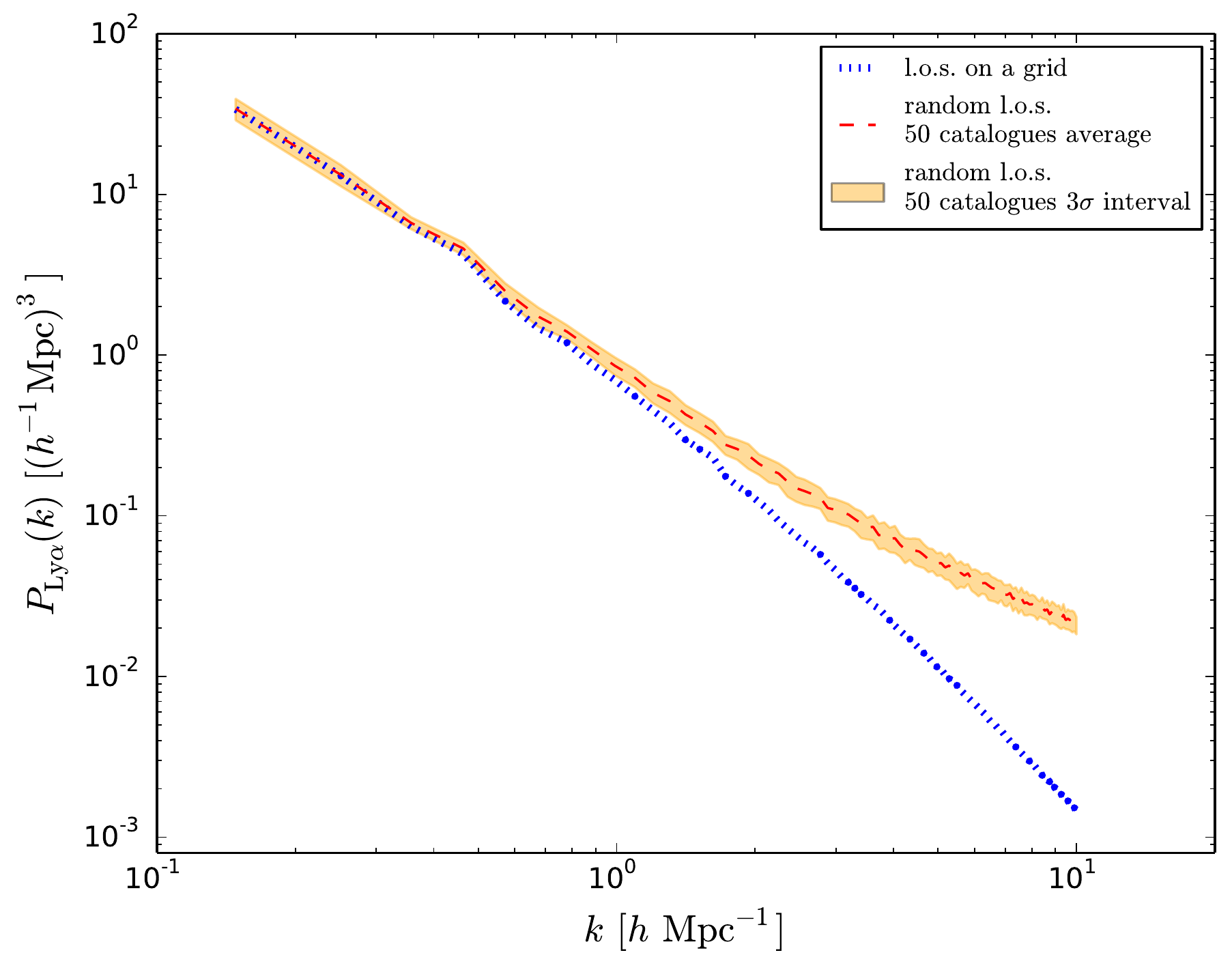}
\caption{3D Ly$\alpha$ flux power spectra $P_{{\rm Ly}\alpha}(k)$ at $z=3$ obtained by placing the skewers into a regular grid (dotted blue) or distributing them randomly (dashed red). For the regular grid we employ $256^2$ spectra while for the randomly distributed skewers we create 50 independent catalogues, each of them containing $256^2$ spectra, and we show the mean of them and the $3\sigma$ variation.}
\label{fig:gridPk}
\end{figure}

The question on the proper number of l.o.s. to use and on the way to place them in the simulation box is related to another central issue. To compute $P_{{\rm Ly}\alpha}(k)$ we need a good coverage of skewers in the box because we discretise the box into cells, make 3D flux density maps and perform discrete Fourier transforms: having cells not crossed by a l.o.s. would be a concern because no flux information is available, hence it will be impossible to assign a field value.

There are several solutions to this problem: 1) compute the 3D Ly$\alpha$ power spectrum masking out the regions not sampled by the mock spectra and 2) working in configuration-space (i.e. compute correlation functions), as in \cite{slosar2009,baoBOSS}. Unfortunately, these solutions are computationally expensive and in the case of 1) technically hard to implement.

The standard practice is to place the skewers into a regular grid, this way the whole box is sampled uniformly, although real quasars are not on a grid. We check what is the effect of regularising the spectra position. In Fig. \ref{fig:gridPk} we plot with a red dashed line the average of 50 Ly$\alpha$ power spectra, each corresponding to a catalogue with $256^2 = 65536$ l.o.s. randomly placed, and with a dotted blue line the $P_{{\rm Ly}\alpha}(k)$ of a catalogue of the same number of l.o.s. but placed on a $256\times256$ regular grid in the $y-z$ plane. The two power spectra converge at large scales, the sub-sampling effect discussed above becomes relevant on small scales.

We thus conclude that our method to compute the 3D Ly$\alpha$ power spectrum is robust, on large-scales, against the different ways of sampling the Ly$\alpha$ field.

\section{Gaussian errors derivation}
\label{app_errors}

Here we derive the equations governing the Gaussian errors of the multipoles of auto- and cross- power spectra.

{\bf Fitting simultaneously monopoles and quadrupoles of the auto-power spectra.} 
If we have measurements of the monopoles and quadrupoles of the auto-power spectra of the two fields that occupy the same volume and we want to fit them simultaneously we need a theoretical model and an estimation of the errors. The theoretical model is given by the Legendre expansion of Eq. \ref{eq:legendre}, that translates to Eqs. \ref{eq:multipoles}-\ref{eq:multipoles2} for the two fields we are considering (dubbed with $\alpha$ and $\beta$ subscripts in what follows). We derive the associated covariance:
\begin{eqnarray}
\sigma^2&(P_{\ell_1,\alpha}&(k_i),P_{\ell_2,\beta}(k_j))=\left \langle \left(\hat{P}_{\ell_1,\alpha}(k_i)-P_{\ell_1,\alpha}(k_i)\right) \left(\hat{P}_{\ell_2,\beta}(k_j)-P_{\ell_2,\beta}(k_j)\right) \right \rangle\nonumber\\
&=&\langle \hat{P}_{\ell_1,\alpha}(k_i)\hat{P}_{\ell_2,\beta}(k_j) \rangle-P_{\ell_1,\alpha}(k_i)P_{\ell_2,\beta}(k_j)\nonumber\\
&=&\frac{(2\ell_1+2)(2\ell_2+1)}{N_k^2}\sum_{l=1}^{N_k}\sum_{m=1}^{N_k}\langle \delta_\alpha(k_{i,l})\delta^*_\alpha(k_{i,l}) \delta_\beta(k_{j,m})\delta^*_\beta(k_{j,m}) \rangle L_{\ell_1}(\mu_l) L_{\ell_2}(\mu_m)\nonumber\\
&&-P_{\ell_1,\alpha}(k_i)P_{\ell_2,\beta}(k_j)\nonumber\\
&=&\frac{(2\ell_1+2)(2\ell_2+1)}{N_k^2}\sum_{l=1}^{N_k}\sum_{m=1}^{N_k}\langle \delta_\alpha(k_{i,l})\delta^*_\beta(k_{j,m}) \rangle \langle \delta^*_\alpha(k_{i,l})\delta_\beta(k_{j,m}) \rangle L_{\ell_1}(\mu_l) L_{\ell_2}(\mu_m) \nonumber\\
&=&\frac{(2\ell_1+2)(2\ell_2+1)}{N_k^2}\sum_{l=1}^{N_k}\sum_{m=1}^{N_k}P^2_{\alpha\beta}(k_{i,l}) L_{\ell_1}(\mu_l) L_{\ell_2}(\mu_m) \delta_{l,m}\delta_{k_i,k_j}\nonumber\\
&=&\frac{(2\ell_1+2)(2\ell_2+1)}{N_k^2}\sum_{l=1}^{N_k}P_{\alpha\beta}^2(k_{i,j})L_{\ell_1}(\mu_l) L_{\ell_2}(\mu_l)\delta_{k_i,k_j}
\end{eqnarray}
where $N_k$ is the number of independent modes in the k-interval $[k,k+dk]$, i.e.
$ N_k = \frac{1}{2}\frac{4\pi k^2dk}{k_F^3} $ where $k_F=2\pi/L$ is the value of the fundamental frequency, with $L$ being the size of the cubic volume and $dk$ is the k-bin size, usually chosen as $dk=k_F$. We notice that in the previous expression we have taken into account that the imaginary part of the cross-power spectrum is 0\footnote{This arises by assuming that $\xi(\vec{r})$ is even, i.e. $\xi(\vec{-r})=\xi(\vec{r})$.}, i.e.
\be
\Im(P_{12}(\vec{k}))=\frac{1}{2}\left( \langle \delta_1\delta^*_2\rangle-\langle \delta_1^*\delta_2\rangle \right)=0
\ee
thus, $\langle\delta_1(\vec{k}_1)\delta^*_2(\vec{k}_2)\rangle=P_{12}(\vec{k})$. In the continuous limit, the above equation can be expressed as
\be
\sigma^2(P_{\ell_1,\alpha}(k_1),P_{\ell_2,\beta}(k_2))=\delta_{k_1,k_2}\frac{(2\ell_1+2)(2\ell_2+1)}{2N_k}\int_{-1}^1 P^2_{\alpha\beta}(k,\mu)L_{\ell_1}(\mu) L_{\ell_2}(\mu)d\mu
\ee
and taking into account that in redshift-space $P_{\alpha\beta}(k,\mu)=b_\alpha b_\beta (1+\beta_\alpha\mu^2)(1+\beta_\beta\mu^2)P_m(k)$ we obtain
\begin{eqnarray}
\sigma^2(P_{0,\alpha}(k_1),P_{0,\beta}(k_2))&=\gamma(k)\bigg[&\frac{1}{9}\beta_\alpha^2\beta_\beta^2 + \frac{2}{7}(\beta_\alpha^2\beta_\beta+\beta_\alpha\beta_\beta^2)+\nonumber\\ && +\frac{1}{5}(\beta_\alpha^2+4\beta_\alpha\beta_\beta+\beta_\beta^2)+\frac{2}{3}(\beta_\alpha+\beta_\beta)+1\bigg]\delta_{k_1,k_2}\nonumber\\
\sigma^2(P_{2,\alpha}(k_1),P_{2,\beta}(k_2))&=5\gamma(k)\bigg[& \frac{415}{1287}\beta_\alpha^2\beta_\beta^2 +\frac{170}{231}(\beta_\alpha^2\beta_\beta + \beta_\alpha\beta_\beta^2) +\nonumber\\  & & + \frac{3}{7} (\beta_\alpha^2 + 4\beta_\alpha\beta_\beta + \beta_\beta^2) +\frac{22}{21}(\beta_\alpha+\beta_\alpha)+1 \bigg]\delta_{k_1,k_2}\nonumber\\
\sigma^2(P_{0,\alpha}(k_1),P_{2,\beta}(k_2))&=\frac{4}{693}\gamma(k)\bigg[& 70\beta_\alpha^2\beta_\beta^2 + 165(\beta_\alpha^2\beta_\beta + \beta_\alpha\beta_\beta^2) +\nonumber\\  && + 99(\beta_\alpha^2+4\beta_\alpha\beta_\beta+ \beta_\beta^2) + 231(\beta_\alpha+\beta_\beta)\bigg]\delta_{k_1,k_2}
\end{eqnarray}
where
\be
\gamma(k)=\frac{b_\alpha^2b_\beta^2P^2_m(k)}{N_k}
\ee

{\bf Fitting simultaneously auto- and cross-power spectra.} 
If we have measurements of monopoles and quadrupoles of the two auto-power spectra and measurements of monopole and quadrupole of their cross-power spectrum and we want to fit all six functions together, we need again a theoretical model and an estimation of the errors. The theoretical model is given again by the expansion in Legendre polinomial as in Eqs. \ref{eq:multipoles}-\ref{eq:multipoles2} and \ref{eq:multicross}-\ref{eq:multicross2}. It follows that the covariance will be given by
\be
\sigma^2(P_{\ell_1,\alpha}(k_1),P_{\ell_2,12}(k_2))=\delta_{k_1,k_2}\frac{(2\ell_1+2)(2\ell_2+1)}{2N_k}\int_{-1}^1 P_\alpha(k,\mu)P_{12}(k,\mu)L_{\ell_1}(\mu) L_{\ell_2}(\mu)d\mu
\ee
and taken into account that $P_\alpha(k,\mu)=b_\alpha^2(1+\beta_\alpha\mu^2)^2P_{\rm m}(k)$ and $P_{12}(k,\mu)=b_1b_2(1+\beta_1\mu^2)(1+\beta_2\mu^2)P_{\rm m}(k)$ we get
\begin{eqnarray}
\sigma^2(P_{0,1}(k_1),P_{0,12}(k_2))&=&\gamma(k)\left[\frac{1}{9}\beta_1^3\beta_2 + \frac{1}{7}(\beta_1^3+3\beta_1^2\beta_2)+\frac{1}{5}(3\beta_1^2+3\beta_1\beta_2)+\frac{1}{3}(3\beta_1+\beta_2)+1\right]\delta_{k_1,k_2}\nonumber\\
\sigma^2(P_{2,1}(k_1),P_{2,12}(k_2))&=&5\gamma(k)\left[ \frac{415}{1287}\beta_1^3\beta_2 + \frac{85}{231}(\beta_1^3 + 3\beta_1^2\beta_2) + \frac{9}{7} (\beta_1^2 + \beta_1\beta_2) +\frac{11}{21}(3\beta_1+\beta_2)+1 \right]\delta_{k_1,k_2}\nonumber\\
\sigma^2(P_{0,1}(k_1),P_{2,12}(k_2))&=&\frac{2}{693}\gamma(k)\left[ 140\beta_1^3\beta_2 + 165(\beta_1^3 + 3\beta_1^2\beta_2) + 594(\beta_1^2+\beta_1\beta_2) + 231(3\beta_1+\beta_2)\right]\delta_{k_1,k_2}\nonumber
\end{eqnarray}
where
\be
\gamma(k)=\frac{b_1^3b_2P^2_m(k)}{N_k}\,.
\ee

\bibliographystyle{JHEP}
\bibliography{Bibliography} 

\end{document}